\documentclass[a4paper,11pt,nofootinbib,floatfix]{revtex4-2}

\usepackage{amssymb,bm,epsf}
\usepackage{setspace}
\usepackage{amsmath}
\usepackage{graphicx}
\usepackage{dcolumn}
\usepackage{color}
\usepackage{accents}
\usepackage{multirow}
\usepackage{tikz}
\usepackage{comment}
\usepackage{subfigure}
\usepackage{soul}
\usepackage[rightcaption]{sidecap}
\sidecaptionvpos{figure}{t}
\usepackage{enumerate}

\usepackage{xcolor}

\usepackage[toc,page]{appendix}
\usepackage[normalem]{ulem}

\usepackage[hyperfootnotes=false]{hyperref}
\hypersetup{
    colorlinks=true,
    linkcolor=red,
    filecolor=magenta,      
    citecolor=blue
}

\newcolumntype{P}[1]{>{\centering\arraybackslash}p{#1}}
\newcolumntype{M}[1]{>{\centering\arraybackslash}m{#1}}

\makeatletter
\@addtoreset{equation}{section}

\makeatother

\begin{document}
\begin{titlepage}

\title{Proper-time functional renormalization in $O(N)$ scalar models coupled to gravity}

\author{Alfio Bonanno}
\email{alfio.bonanno@inaf.it}
\affiliation{
INAF Osservatorio Astrofisico di Catania, Via S.Sofia 78, 95123 Catania ITALY
\\
and 
INFN, Sezione di Catania, Italy
}
\author{Emiliano Glaviano}
\email{emiliano.glaviano@inaf.it}
\affiliation{
INAF Osservatorio Astrofisico di Catania, Via S.Sofia 78, 95123 Catania ITALY
\\
and 
INFN, Sezione di Catania, Italy
}
\author{Gian Paolo Vacca}
\email{vacca@bo.infn.it}
\affiliation{INFN, Sezione di Bologna,
via Irnerio 46, I-40126 Bologna}

\begin{abstract}
We focus on the use of the functional Wilsonian renormalization group framework characterized by a proper time regulator and test its use in the search of the scaling solutions and the critical properties of an $O(N)$-invariant scalar field multiplet coupled to gravity in $d=4$ and $d=3$ dimensions. We employ the same background-fluctuation splitting and gauge fixing procedure, already adopted in a previous study based, instead, on the effective average action framework and a similar truncation of the effective action. Our main goal is to compare the results for the scaling solutions and some of the associated critical exponents. In this analysis, performed in a different framework, most of the picture previously uncovered is confirmed both at qualitative and quantitative level. There are, neverthelss, few differences both at finite $N$ and in its large value limit, depending also on the schemes which in both frameworks are called "improved".

\end{abstract} 

\maketitle

\end{titlepage}
\newpage
\setcounter{page}{2}

%%%%%%%%%%%%%%%%%%%%%%%
\section{Introduction}
%%%%%%%%%%%%%%%%%%%%%%%

\noindent 
An important development in modeling fundamental quantum physical phenomena but also quantum and statistical effective critical physics was originated by the developments in the analysis of Quantum (and statistical) Field Theory. An important breakthrough in this direction is related to the comprehensive view obtained within the renormalization group paradigm. The idea born in perturbative studies and culminating in~\cite{PhysRev.95.1300} (Gellmann-low) was completely reformulated by K. Wilson~\cite{PhysRev.140.B445,PhysRevD.2.1438, PhysRevLett.28.240,PhysRevB.4.3174,Wilson:1973jj}, following Kadanoff \cite{PhysicsPhysiqueFizika.2.263}, when several tools and techniques were developed to tackle non-perturbative analysis. Later on analytical techniques were developed to study the renormalization group flow of various types of effective actions. 

In general, all the so-called Wilsonian approaches are based on a two-step procedure: a coarse-graining of the degrees of freedom followed by a spacetime rescaling, controlled by a change of scale. When the latter is infinitesimal, one may formally construct exact integral-differential functional equations that in principle provide, if solved, an alternative way to construct the generating functionals which give access to the physical observables. This is the essence of the Wilsonian functional renormalization group (FRG) approach.

In this framework, two classes of effective actions received attention. One is the so-called Wilsonian action, whose RG flow is controlled by a UV scale which sets the energy scale above which all fluctuations in the functional integral are integrated, This action can be inserted into a suitable path integral over the remaining fluctuations to obtain the full generating function of the theory of interest. Two specific functional integro differential RG flow equations were constructed by Wegner-Houghton~\cite{PhysRevA.8.401} and Polchinski~\cite{Polchinski:1983gv}. Generalizations have been discussed in several works; see for example~\cite{Rosten:2010vm}. Recently, a special class of RG flows for the Wilsonian action based on a Schwinger proper-time (PT) regulator \cite{Fock:1937dy,PhysRev.82.664} for the coarse-graining procedure was discussed~\cite{deAlwis:2017ysy, Bonanno:2019ukb,Bonanno:2025tfj} and in particular in~\cite{Bonanno:2019ukb} it is argued how RG flow equations in some PT scheme might be interpreted as exact in the UV-regulated Wilsonian RG sense.

The PT RG flow equations were previously used in the literature~ \cite{Bonanno:2000yp,Mazza:2001bp,Liao:1994fp,Liao:1995nm,Schaefer:1999em,Schaefer:2001cn,Bohr:2000gp,Zappala:2001nv,Litim:2010tt, Bonanno:2022edf} but not obtained within a functional derivation.
The other popular class of Wilsonian RG flow analysis is based on the study of the 1PI effective average action (EAA), related by a kind of Legendre transform to the Polchinski Wilsonian action. Its RG flow is given by the Wetterich-Morris equation~\cite{Morris:1993qb,WETTERICH199390, Ellwanger:1993mw} (look also at \cite{book,Reuter_Saueressig_2019} for a pedagogical introduction and \cite{Reuter:1996cp,Niedermaier:2006ns,Reuter_2012,Eichhorn:2018yfc,Dupuis:2020fhh,Eichhorn:2017egq,Bonanno:2020bil} for reviews and applications to quantum gravity), which has better convergence properties compared to the ones following from the Polchinski equation. Because of that, they are heavily used in the study of critical phenomena. It is by now well understood that proper-time (PT) flow equations are, in general, not exact when formulated for an IR-regulated 1PI effective action~\cite{Litim:2001ky}. Nevertheless, PT flows have recently been advocated as a useful approximate scheme in this context~\cite{wetterich2024simplifiedfunctionalflowequation}.

Fixed points of the RG flow, where scale invariance is realized, are associated to the critical theories, which describe the universal behavior of physical systems. In particular infrared attractive critical theories describe the large distance universal behavior, shared by physically possible very different microscopic models falling in the same universality class. The universal properties are extracted by linear deformations around the critical theory studying the eigenfunctions and the universal eigenvalues of the associated linearized flow operator. Looking at the ultraviolet (UV) fixed points a generalization of the concept of renormalizability called asymptotic safety was pushed forward for nontrivial critical interacting theories, with a finite number of attractive deformations, case which goes beyond the asymptotic freedom property determined by a trivial gaussian UV fixed point. Asymptotic safety was introduced by S. Weinberg in the discussion of the possibility that gravitational interactions might be renormalizable in this sense, since the existence of such a fixed point was originally discovered in~\cite{Weinberg:1976xy,Weinberg:1980gg} for gravity, as a quantum field theory of the metric field which is a gauge theory with diffeomorphism invariance, in $d=2+\varepsilon$ dimensions.

The determination of a non-Gaussian fixed point for the theory of gravitation, first studied intensively in its basic Einstein-Hilbert formulation (truncation), was considered a crucial step in understanding both the non-perturbative renormalizability of the underlying field theory and, more generally, a possible path to clarify the quantum nature of gravity. Results on its existence have been determined for pure gravity \cite{Reuter:2001ag,Lauscher:2002sq,deBerredo-Peixoto:2004cjk,Ohta:2013uca,Ohta:2015zwa,Codello:2006in,Machado:2007ea,Codello:2007bd,Benedetti:2012dx,Falls:2013bv,Benedetti:2013jk,Falls:2014tra,Falls:2017lst,Falls:2018ylp,DeBrito:2018hur,Ohta:2015efa,Ohta:2015fcu,Gies:2016con,Falls:2016msz,Christiansen:2016sjn,Denz:2016qks,Falls:2020qhj,Sen:2021ffc,Kawai:2023rgy,Kawai:2024rau,Benedetti:2009rx,Benedetti:2009gn} and for gravity coupled with matter \cite{Percacci:2003jz,Narain:2009fy, Narain:2009gb,Vacca:2010mj,Eichhorn:2012va,Dona:2013qba,Labus:2015ska,Oda:2015sma,Hamada:2017rvn,Eichhorn:2017als,Pawlowski:2018ixd,Wetterich:2019zdo,Meibohm:2015twa,Christiansen:2017cxa,Christiansen:2017bsy,Ohta:2021bkc,Pastor-Gutierrez:2022nki,Niedermaier:2006wt,Niedermaier:2006ns,Percacci:2007sz,Kotlarski:2023mmr}
%~\cite{Reuter:1996cp,Falkenberg:1996bq,Lauscher:2001ya,Reuter:2001ag,Codello:2006in,Codello:2007bd,Codello:2008vh,Nink:2014yya,Gies:2015tca,DeBrito:2018hur,Bonanno:2004sy} 
were given for the $d=4$ case mainly studying the Wetterich-Morris equation, even if, because of many approximations taken, a complete understanding of the ultraviolet (UV) critical manifold's structure in four dimensions remains an active area of investigation. Even with more general gravitational truncations, namely with higher order curvature terms in the effective action and with the presence of different kind of matter fields, the presence of a fixed point has been always found (see \cite{Bonanno:2020bil} for a recent discussion.) 

In this work we analyze the properties of a set of scalar fields with $O(N)$ internal symmetry in the vector representation coupled to gravity studying the RG flow of the above mentioned Wilsonian action with a family of PT regulators. The main goal is to compare the results previously obtained by one of us~\cite{Percacci:2015wwa, Labus:2015ska} in the context of the RG flow of the 1PI effective average action. In that analysis a specific parameterization of the metric was used (exponential) combined with the so-called "physical" gauge fixing in order to obtain RG flow equations as simpler as possible for a truncation of the action depending on two functions to be determined at the fixed points.

One of the motivations to perform this analysis is due to the fact that it has been observed that PT RG flows sometimes provide a remarkably accuracy in determining observables (even at low orders in the derivative expansion), while preserving important symmetries of the action. This symmetry-preserving property, well-established for gauge theories \cite{Liao:1995nm} through the gauge-invariant proper time regulator introduced in \cite{Schwinger:1951nm}, makes it particularly interesting for quantum gravity applications.

As recently discussed in \cite{Falls:2024noj}, the PT flow can also be viewed as a variant of dimensional regularization that handles poles appearing in all even dimensions $d$. This approach, combined with field redefinitions to eliminate off-shell contributions to RG equations in the spirit of the essential renormalization group \cite{Falls:2017cze, Baldazzi:2021ydj}, focuses specifically on the flow of couplings relevant to physical observables. Notably, this framework has demonstrated parameterization independence for the Newton's constant beta function to all orders in the scalar curvature.

Let us recall here the general form of RG flow equation for the Wilsonian action  $S_\Lambda$ with a PT regulator which we shall use:
\begin{equation}
\label{PTFE}
\Lambda\partial_\Lambda S_\Lambda  [\phi] = \frac{1}{2}\int_{0}^{\infty}\frac{ds}{s}r\left(s,\Lambda^2 Z_\Lambda\right)\mathrm{Tr}\left[e^{-sS_\Lambda^{(2)}}\right]\,,
\end{equation}
where $S_\Lambda^{(2)}$ is the Hessian of the theory, $r(s,\Lambda^2 Z_\Lambda)$ is a cutoff function, and $Z_\Lambda$ is the wavefunction renormalization associated with the fields of the theory.

Before proceeding, we clarify the logical status of the proper-time RG flow equation~\eqref{PTFE} used in this work. As it has been discussed in~\cite{Bonanno:2019ukb} we consider the possibility that such RG scheme can be derived exactly in a UV-regulated Wilsonian sense~\footnote{ We stress that a particular case of Eq.~\eqref{PTFE} was derived by de Alwis ~\cite{deAlwis:2017ysy} by introducing a specific family of regulatization schemes under the assumption that a functional determinant could be written using a Schwinger proper time representation in terms of the Wilsonian regularized action. This is specifically what was formally avoided in the approach presented in~\cite{Bonanno:2019ukb}.}. To the best of our knowledge this is a possibility which is not ruled out and still to be fully analyzed since it might have a non trivial realization. In the Appendix~\ref{appendix0} we provide for the reader's convenience a short review on a general, slightly formal, approach used to construct the RG flow of a UV-regulated Wilsonian action and then write the functional equations which one may solve to construct a PT scheme at two different levels.

In this work we {\it assume} that at least one physically admissible solution for the quantities (see Appendix~\ref{appendix0}) $\psi_x^\Lambda$ or $b^\Lambda$ exists, with reasonable properties and without overly severe non-localities. No attempt is done to make a construction since this is completely out of the scope of this work and certainly deserves a separate study.

Following \cite{Bonanno:2019ukb}, we use the spectrally adjusted cutoff function:
\begin{equation}
\label{cutoff}
r(s,\Lambda^2 Z_\Lambda) = \left(2 + \epsilon\frac{\Lambda \partial_\Lambda Z_\Lambda}{Z_\Lambda}\right)\frac{(s\, m\, \Lambda^2 Z_\Lambda)^m}{\Gamma(m)}e^{-s\, m\, \Lambda^2 Z_\Lambda}\,,
\end{equation}
where $m$ is an arbitrary positive real number that controls the behavior of the cutoff family $r(s,\Lambda^2 Z_\Lambda)$ in the interpolating region. The dependence in the product $s Z_\Lambda$ in the power and exponential is dictated by the requirement of performing a suitable rescaling after the coarse-graining in the  RG step defining the Wilosonian flow. $\epsilon$ distinguishes between two types of cutoff functions: type B ($\epsilon=1$) and type C ($\epsilon=0$). In the $m\to \infty$ limit the RG flow equation simplifies to
\begin{equation}
\label{PTFEexp}
\Lambda \partial_\Lambda S_\Lambda  [\phi] = {\rm Tr}\left[ \left( 1+ \frac{\epsilon}{2}\frac{\Lambda \partial_\Lambda Z_\Lambda}{Z_\Lambda}\right)
e^{-\frac{S_\Lambda^{(2)}}{\Lambda^2 Z_\Lambda}}\right]\,.
\end{equation}

In particular our work aims to explore the critical properties the PT flow for $O(N)$ scalar theories in gravitational backgrounds in $d=3$ and $d=4$ dimensions for various choice of the regulator and of the coarse graining scheme.  In $d=3$ and for $N=1$ this approach allows to investigate the properties of a gravitationally dressed Wilson-Fisher fixed point, a deformations of the Ising universality class, with other values of $N$ being also of interest.

In Section~\ref{sect:2} the specific model to be studied is presented and the system of flow equations is given. In section \ref{secAS} and \ref{secNinf} we describe the analytical set of fixed-point solutions, followed by their critical properties. In section \ref{secGWF} we present the numerical gravitational-dressed Wilson-Fisher fixed-point solutions,  and in section \ref{seccritWF} their critical properties. In section \ref{conc} we present the conclusions. Finally there are five appendices, the first \ref{appendix0} is a short general review of UV-regulated wilsonian flow equations, the other two,~\ref{appendix1} and~\ref{appendix2}, cover the steps for the derivation of the RG equations for the selected truncation of the effective action, while the latter two,~\ref{appendix3} and~\ref{appendix4}, are devoted to discuss some detail of the numerical analysis. 

\section{The Flow Equations}\label{sect:2}
\noindent We study the most general theory coupling Einstein gravity non-minimally to $N$ scalar fields:
\begin{equation}
\label{theory}
S\left[g,\phi_1,\ldots,\phi_N\right] = \int{d^dx\sqrt{g}\left(\frac{1}{16\pi G}\left(-R + 2\lambda\right) + \frac{1}{2}\sum_{i=1}^{N}\phi_i (-\Box)\phi^i + B\left(\rho\right)R + V\left(\rho\right)\right)},
\end{equation}
where $-\Box$ is the Laplacian operator acting on the $i$-th scalar field $\phi_i$. We assume that the $N$ scalar fields form a multiplet $\phi = (\phi_1,...,\phi_N)$ transforming  as a fundamental representation of the $O(N)$ group. The variable $\rho$ is defined as
\begin{equation}
\rho = \frac{1}{2}\sum_{a=1}^{N}\phi^a\phi^a.
\end{equation}
By introducing
\begin{equation}
\label{sosFU}
F\left(\rho\right) = \frac{1}{16\pi G} - B\left(\rho\right), \quad U\left(\rho\right) = V\left(\rho\right) + \frac{\lambda}{8\pi G},
\end{equation}
we obtain a more concise form of the action:
\begin{equation}
\label{theorysem}
S\left[g,\phi_1,...,\phi_N\right] = \int{d^dx\sqrt{g}\left(-F\left(\rho\right)R + \frac{1}{2}\sum_{i=1}^{N}\phi_i (-\Box)\phi^i + U\left(\rho\right)\right)}.
\end{equation}
In this form, the action can be viewed as a generalization of the Local Potential Approximation (LPA) that includes two derivatives of the metric.

While Eq.~(\ref{theorysem}) represents a classical action, quantum effects induce running couplings for $F$ and $U$. Our goal is to study the Wilsonian flow of these running couplings, meaning that Eq.~(\ref{theorysem}) becomes a Wilsonian action. In general the Wilsonian action is a complicated non local functional but, as usually done, we shall consider an approximated form, a truncation, which will depend on two flowing functions $F_\Lambda$ and $U_\Lambda$:
\begin{equation}
\label{theorysemk}
S_\Lambda \left[g,\phi_1,...,\phi_N\right] = \int{d^dx\sqrt{g}\left(-F_\Lambda\left(\rho\right)R + \frac{1}{2}\sum_{i=1}^{N}\phi_i(-\Box)\phi^i + U_\Lambda\left(\rho\right)\right)}.
\end{equation}
where $\Lambda$ is the Wilsonian UV cutoff.

In previous works \cite{Narain:2009fy, Percacci:2015wwa, Labus:2015ska, Henz:2016aoh}, the Wetterich equation for the one-particle irreducible (1PI) infrared-regulated effective action was used to study the running of $F$ and $U$. Here, we instead employ the proper time flow equation, which describes the flow of the Wilsonian action, as anticipated in the introduction. 

For our effective action in Eq.~(\ref{theorysemk}), the left-hand side of the flow equation becomes
\begin{equation}
\label{LHSPT}
\Lambda\partial_\Lambda S_\Lambda = \int{d^dx\sqrt{g}\left(-\Lambda\partial_\Lambda F_\Lambda(\rho)R + \Lambda\partial_\Lambda U_\Lambda(\rho)\right)}.
\end{equation}
Consequently, to extract the running couplings of $F$ and $U$, the right-hand side of Eq.~(\ref{PTFE}) must be projected onto terms proportional to $R^0$ and $R^1$. This projection can be performed using the background field method followed by a heat kernel expansion \cite{Barvinsky:1990up, Barvinsky1987BeyondTS, Barvinsky:1990uq, Barvinsky:1993en, Avramidi:2000bm}. For the scalar fields, we employ the standard linear split $\phi_a = \bar{\phi}_a + \Phi_a(x)$, while for the gravitational field we use the exponential split $g_{\mu\nu} = \bar{g}_{\mu\rho}(e^h)^\rho_{\ \nu}$, in order to compare with a previous analysis \cite{Narain:2009fy, Percacci:2015wwa, Labus:2015ska} done using the effective average action formalism. Here, $\bar{\phi}_a$ and $\bar{g}_{\mu\rho}$ are fixed but arbitrary background fields, and $\Phi_a(x)$ and $h_{\mu\nu}$ are the fluctuation fields. The technical details for deriving the flow equations are provided in Appendices~\ref{appendix1} and~\ref{appendix2}. 

Using the results from Eqs.~(\ref{FlowU}) and (\ref{FlowF}) in the appendices, and defining the dimensionless rescaled variables $x = \alpha^{1/2} \Lambda^{2-d}\rho$, $u_\Lambda(x) =\alpha \Lambda^{-d} U_\Lambda(\rho)$, and $f_\Lambda(x) = \alpha \Lambda^{2-d} F_\Lambda(\rho)$, where $\alpha=\frac{m^{d/2}\Gamma\left(m-\frac{d}{2}\right)}{4(4\pi)^{d/2}\Gamma(m)}$, we obtain the dimensionless flow equations for $u_\Lambda$ and $f_\Lambda$:
\begin{equation}\begin{split}
\label{betau}
&\dot{u}=-d u +(d-2) x u'+2 (d-3) d+4 (N-1) \left(1+\frac{u'}{m}\right)^{\frac{d}{2}-m}+\epsilon (d-1) d  \left((d-2)\left(1-x \frac{f'}{f}\right)+\frac{\dot{f}}{f}\right)
%+
\\
& +4\left(1+\frac{\epsilon 
   \left(\frac{1}{2} (d-2) \left(\frac{x f'}{f}-\frac{2 x f''}{f'}-1\right)+\frac{\dot f'}{f'}-\frac{\dot f}{2 f}\right)}{1+\frac{(d-2) f}{4 (d-1) x \left(f'\right)^2}}\right)\left(1+\frac{2 x u''+u'}{m\left(1+\frac{4 (d-1) x \left(f'\right)^2}{(d-2) f}\right)}\right)^{\frac{d}{2}-m}\,,
\end{split}\end{equation}

\begin{equation}\begin{split}
\label{betaf}
&\dot{f}=(2-d) \left(f-x f'\right)+\frac{1}{3}\left(\frac{d}{2m}-1\right)\left(d^2-3d+36\right)+4\left(\frac{d}{2m}-1\right)\left(N-1\right)\left(\frac{1}{6}+f^\prime\right)\left(1+\frac{u^\prime}{m}\right)^{\frac{d}{2}-m-1}
%+
\\
&+ \epsilon\frac{1}{6}\left(\frac{d}{2m}-1\right)\left(d^2-d-\frac{24}{d}-24\right) \left((d-2) \left(1-x \frac{f'}{f}\right)+\frac{\dot f}{f}\right)+\\
   &\resizebox{1.05\hsize}{!}{$+\frac{2}{3}\left(\frac{d}{2m}-1\right) \left(1+\frac{6 \left(\frac{4 x \left(f'\right)^2}{(d-2) f}+2 x f''+f'\right)}{1+\frac{4 (d-1) x \left(f'\right)^2}{(d-2)
   f}}\right)
   \left(1+\frac{\epsilon \left(\frac{1}{2}(d-2) \left(\frac{x f'}{f}-\frac{2 x f''}{f'}-1\right)+\frac{\dot f'}{f'}-\frac{\dot f}{2 f}\right)}{1+\frac{(d-2) f}{4 (d-1) x
   \left(f'\right)^2}}\right)\left(1+\frac{2 x u''+u'}{m\left(1+\frac{4 (d-1) x \left(f'\right)^2}{(d-2) f}\right)}\right)^{\frac{d}{2}-m-1}$} \!\!\!.
\end{split}\end{equation}
A "$\prime$" indicates the derivative with respect to $x$, whereas a dot the derivate with respect to the RG time $t=\ln\left(\Lambda/\Lambda_0\right)$, where $\Lambda_0$ is an arbitrary normalization scale. Due to the properties of the regulator that enter in the domain of $\alpha$, these equations are defined for $m-d/2>0$. 

On setting $\dot{u}=0$, $\dot{f}=0$ and $\dot{f'}=0$ one obtains the fixed-point equations. At the fixed point the equations turn into an ordinary non-linear second-order system of differential equations. The solutions $u_*(x)$ and $f_*(x)$ are labeled with four independent parameters that have to be fixed by boundary conditions. The structure of scaling solutions constrain the boundary conditions and so the free parameters; consequently, there are at most a discrete set of acceptable fixed-point solutions. In the next sections, we describe for various values of the number of fields and regulator type the set of fixed-point solutions. 

\section{Analytical scaling solutions and their properties}\label{secAS}
%\subsection{The scaling solutions}
\noindent In this section, we present analytical scaling solutions for both the $\epsilon = 0$ (C-type cutoff) and $\epsilon = 1$  (B-type cutoff) cases.
For $\epsilon = 0$, the PT flow equation can be derived as a renormalization group (RG) improvement of a standard one-loop calculation \cite{Bonanno:2004sy}. This limit corresponds to the ``unimproved" Wetterich flow equation discussed in \cite{Percacci:2015wwa}. In contrast, the $\epsilon = 1$ case, while interpretable as a particular coarse-grained flow \cite{Bonanno:2019ukb}, lacks an immediate physical interpretation. It corresponds to the "improved" cutoff introduced in \cite{Percacci:2015wwa}.

The following fixed-point solutions exist for any value of the cutoff and dimension (we will consider $d>2$ in the following):
\begin{equation}\begin{split}
\label{GFP}
&u_*=2d+\frac{4N}{d}-6+\left(d^2-3d+2\right)\epsilon\\
&f_*=\frac{(d-2m)(d^2-3d-36+2N)}{6(d-2)m}+\frac{(d^3-d^2-24d-24)(d-2m)\epsilon}{12dm}
\end{split}\end{equation}
 The value of $u_*$ does not depend on the cutoff parameter $m$. The sign of $f_*$ depends on the number of fields and $d$. In particular for
 \begin{equation}
N>N_c= \frac{1}{2} \left(-d^2+3 d+36\right)+\frac{1}{4} \left(-d^3+3 d^2+22
   d-\frac{48}{d}-24\right) \epsilon
 \end{equation}
$f_*$ takes always a negative value. In the $\epsilon=0$ case we also find the additional solution
\begin{equation}\begin{split}
\label{scal1}
&u_*(x)=\frac{4 N}{d}+2 d-6, \quad \quad f_*(x)=\frac{(d-2 m) \left(d^3-4 d^2+d (2 N-33)+10 N+36\right)}{6 (d-2) (d-1) m}+\frac{x}{d-1}
\end{split}\end{equation}
where value of $u_*$ is the same of the above fixed point but now $f_\ast$ has a linear dependence on the field $x$ (see also \cite{Labus:2015ska}).
%The value of the coefficient of $x$ is the same of \cite{Labus:2015ska}.  
As before the sign of $f_*(0)$ depends on the number of matter fields and in this case 
\begin{equation}
\label{NCs1}
N_c=-\frac{d^3-4 d^2-33 d+36}{2 (d+5)}
\end{equation}
above this value $f_*(0)<0$. 

If we set $f_0=0$ we find a solution with a constant $u_*$ and a linear $f_*$. In this case one notes that $x f'/f=1$ so that the dependence on $\epsilon$ drops in both the fixed point equations for $u$ and $f$. As a consequence there is no difference between the two schemes based on the B- and the C-type cutoffs. The flow equations are quadratic in $f_*$ and admit two real solutions. We show the results only for $d=3$ and $d=4$ because the expressions for a general $d$ are quite long and not particularly illuminating. In $d=3$ we get
\begin{equation}\begin{split}
\label{gfp1}
&u_*=\frac{4 N}{3}, \quad \quad f_*(x)=\frac{60-7N\pm \sqrt{N^2+72 N+2736}}{48 (N-1)} x
\end{split}\end{equation}
and in $d=4$ 
\begin{equation}\begin{split}
\label{gfp2}
&u_*=2+N,\quad \quad f_*(x)=\frac{7-N\pm\sqrt{3} \sqrt{N+11}}{6 (N-1)}x
\end{split}\end{equation}
again the value of $u_*$ is the gaussian value, here for $d=3$ and $d=4$. These solutions do not have any dependence on $m$. Of the two solutions, only the one with the minus sign has a regular limit $N\to1$. This solution always yields a negative $f_*$. On the contrary in the solution with the plus sign there is a critical value above which $f_*$ becomes negative. The critical values are $N=18$ and $N=16$ for $d=3$ and $d=4$ respectively. The same behavior but with different critical values of $N$ was observed in \cite{Labus:2015ska} for the analysis based on the solutions of the effective average action (1PI) RG flow.

\subsection{Stability analysis for the analytical scaling solutions}
In order to discuss the critical properties let us consider the linear perturbations
\begin{equation}
\label{anslin}
u_\Lambda(x)=u_*(x)+\delta u (x)\left(\frac{\Lambda_0}{\Lambda}\right)^{\theta}, \quad \quad f_\Lambda(x)=f_*(x)+\delta f (x)\left(\frac{\Lambda_0}{\Lambda}\right)^{\theta}
\end{equation}
where $\theta$ is a critical exponent so that for $\theta>0$ the corresponding eigenoperator is relevant. The linearized equations around the gaussian fixed-point eq. (\ref{GFP}) read
\begin{equation}\begin{split}
&\left(8-\frac{4d}{m}\right)\delta u^{\prime\prime}+\left(N\left(4-\frac{2d}{m}\right)\frac{1}{x}+2-d\right)\delta u^\prime+\left(d-\theta\right)\frac{{\delta u}}{x}+\\
&+\frac{12\left(d-2\right)\left(d-1\right)d^2\lambda m\epsilon}{\left(d-2m\right)\left(d^4\epsilon+d^3\left(2-3\epsilon\right)-2d^2\left(11\epsilon+3\right)+4d\left(n+6\left(\epsilon-3\right)\right)+48\epsilon\right)}\left(\left(d-2\right)\delta f^\prime-\lambda\frac{{\delta f}}{x}\right)=0\\
&\left(8-\frac{4d}{m}\right)\delta f^{\prime\prime}+\left(N\left(4-\frac{2d}{m}\right)\frac{1}{x}-\frac{2\left(d-2\right)d\left(d^2-3d+2N-36\right)}{d^4\epsilon+d^3\left(2-3\epsilon\right)-2d^2\left(11\epsilon+3\right)+4d\left(N+6\left(\epsilon-3\right)\right)+48\epsilon}\right)\delta f^\prime+\\
&+\left(d-\theta-2+\frac{\left(d-2\right)\left(d^3-d^2-24d-24\right)\theta\epsilon}{d^4\epsilon+d^3\left(2-3\epsilon\right)-2d^2\left(11\epsilon+3\right)+4d\left(N+6\left(\epsilon-3\right)\right)+48\epsilon}\right)\frac{{\delta f}}{x}-\\
&-\frac{\left(d-2m\right)\left(d-2\left(m+1\right)\right)\delta u^{\prime\prime}}{3m^2}-\frac{N\left(d^2-2d\left(2m+1\right)+4m\left(m+1\right)\right)}{6m^2}\frac{\delta u^\prime}{x}=0
\end{split}\end{equation}
These equations can be studied analytically, in particular we can obtain the values of critical exponents analytically. We shall restrict ourselves to the analysis of deformations with respect the $O(N)$ symmetry. Replacing the Frobenius ansatz ${\delta u}\left(x\right)=\sum_{i=0}^{\infty}{u_i x^i}$, ${\delta f}\left(x\right)=\sum_{i=0}^{\infty}{f_i x^i}$ and requiring that no exponential term is contained in the solutions leads to a truncation of the two series to some $i_{max}$ value and then to a quantization condition for the eigenvalues: 
\begin{equation}
\begin{split}
\label{CritExpGFP}
&\theta=-2+d-\left(d-2\right)j+\frac{\left(d-2\right)^2\left(d^3-d^2-24d-24\right)\epsilon}{2d\left(d^2-3d-36+2N\right)}\\
&\theta=d-\left(d-2\right)j
\end{split}\end{equation}
with $i_{max}=j=0,1,\ldots$. The second set is the standard result of scalar field theory, as expected at the gaussian fixed point the scalar potential eigenoperators enjoy a classical scaling. In the C-type cutoff, where $\epsilon=0$, the two sets of critical exponents coincide \footnote[1]{The first set yields ${\delta u}=0$ and $\delta f\ne0$. This solution is contained in the second set as a particular case.}. On the contrary, in the B-type cutoff, where $\epsilon=1$, there are two discrete and independent sets of critical exponents. The first set of critical exponents exists only if $N\ne18-\frac{1}{2}(d-3)d$.

The values of $\theta$ do not depend on $m$ but the solutions do. The general expressions are quite long, we show and discuss only the solutions in $d=4$. For $j=0,1,2$  with $m=\frac{d}{2}+1$, apart from a global multiplicative constant, we get
\begin{equation*}\begin{split}
&\theta=2+\frac{18 \epsilon }{16-N}:\\
&\delta u= \frac{108 \epsilon }{N+9 \epsilon -16},\quad \quad \delta f=1\\
&\theta = \frac{18 \epsilon }{16-N}:\\
&\delta u= \frac{108 \epsilon 
   \left(N^2-N (9 \epsilon +32)+27 \epsilon ^2+144 \epsilon +256\right)}{(N-9 \epsilon -16) (N-3 \epsilon -16) (2 N+9 \epsilon
   -32)}-\frac{162 (N-16)  \epsilon }{N (N-9 \epsilon -16) (N-3 \epsilon -16)}x,\\
&\delta f=1-\frac{3 (N-16)(N+9 \epsilon -16)}{2 N (N-9 \epsilon -16) (N-3 \epsilon
   -16)}x\\
   &\theta=-2+ \frac{18\epsilon }{16-N}:\\
&\resizebox{1.05\hsize}{!}{$\delta u=-\frac{324 \epsilon ^2
   \left(-N^3+3 N^2 (9 \epsilon +16)-3 N \left(15 \epsilon ^2+288 \epsilon +256\right)+81 \epsilon ^3+720 \epsilon ^2+6912
   \epsilon +4096\right)}{(N-9 \epsilon -16) (N+3 \epsilon -16) \left(-2 N^3+3 N^2 (9 \epsilon +32)-6 N \left(9 \epsilon
   ^2+144 \epsilon +256\right)+81 \epsilon ^3+864 \epsilon ^2+6912 \epsilon +8192\right)}+$}\\
   &+\frac{324 (N-16) x \epsilon  \left(N^2-N (9
   \epsilon +32)+27 \epsilon ^2+144 \epsilon +256\right)}{N (N-9 \epsilon -16) \left(-2 N^3+3 N^2 (9 \epsilon +32)-6 N \left(9
   \epsilon ^2+144 \epsilon +256\right)+81 \epsilon ^3+864 \epsilon ^2+6912 \epsilon +8192\right)}+\\
&-\frac{243 (N-16)^2 x^2
   \epsilon  (2 N+9 \epsilon -32)}{N (N+2) (N-9 \epsilon -16) \left(-2 N^3+3 N^2 (9 \epsilon +32)-6 N \left(9 \epsilon ^2+144
   \epsilon +256\right)+81 \epsilon ^3+864 \epsilon ^2+6912 \epsilon +8192\right)},
\end{split}\end{equation*}
\begin{equation}\begin{split} 
\label{ein1GFP}
&\delta f=1+\frac{3 (N-16) x \left(2 N^2+N (3 \epsilon -64)-27
   \epsilon ^2-48 \epsilon +512\right)}{N \left(-2 N^3+3 N^2 (9 \epsilon +32)-6 N \left(9 \epsilon ^2+144 \epsilon
   +256\right)+81 \epsilon ^3+864 \epsilon ^2+6912 \epsilon +8192\right)}\\
   &-\frac{9 (N-16)^2 x^2 \left(2 N^2+N (27 \epsilon -64)+81 \epsilon ^2-432 \epsilon
   +512\right)}{4 N (N+2) (N-9 \epsilon -16) \left(-2 N^3+3 N^2 (9 \epsilon +32)-6 N \left(9 \epsilon ^2+144 \epsilon
   +256\right)+81 \epsilon ^3+864 \epsilon ^2+6912 \epsilon +8192\right)}
\end{split}\end{equation}
for the first set of $\theta$ and
\begin{equation}\begin{split}
\label{ein2GFP}
&\theta=4,\quad \quad \delta u=1,\quad \quad \delta f=0\\
&\theta=2,\quad \quad \delta u=1-\frac{9 x}{14 N},\quad \quad \delta f= \frac{N-9 \epsilon -16}{189 \epsilon }\\
&\theta=0, \quad \quad \delta u=1-\frac{3
   x}{N}+\frac{27 x^2}{28 N (N+2)},\quad \quad \delta f= \frac{2
   (N-30 \epsilon -16)}{189 \epsilon }+\frac{-N+9 \epsilon +16}{63 N \epsilon }x
\end{split}\end{equation}
for the second set. The second set of critical exponents gives in both cutoffs two relevant directions. In the B-type cutoff in the first set, the number of relevant directions is determined by the sign of $\frac{18}{16-N}$. If $N>16$ the only relevant direction can be contained in $\theta=2+\frac{18}{16-N}$. This critical exponent is positive only if $N>25$. If $N<16$ there is a critical value $j_c$ where $\theta=0$. This critical value is $j_c=\frac{9}{16-N}$. The number of relevant directions is given by the integral part of $j_c+1$. For $N=1$ with the first set we obtain two relevant directions. In total, there are four relevant directions. Four relevant directions and two sets of critical exponents have been also found in \cite{Percacci:2015wwa} with the improved Wetterich-Morris equation giving the flow of the 1PI effective avarage action. We find the same values for the second set of $\theta$ but different values for the first set. The eigenvectors share the same structure and only the coefficients are different.

The linearized equations for eq.~(\ref{scal1}) for generic $d$ with $m=d/2+1$ are given by
\begin{equation}\begin{split}
&(d-\theta ) \delta u+\\
&\resizebox{1.05\hsize}{!}{$+\frac{-4 N \left(d^3 (x-4)+d^2 (11 x+16)+4 d (5 x+33)+4 (x-36)\right)-(d+2) x
   \left(2 d^4-3 d^3 (x+4)-2 d^2 (3 x+25)+12 d (x+17)+24 (x-10)\right)+32 (d+5) N^2}{(d+2)
   \left(2 d^3-d^2 (3 x+8)+2 d (2 N-6 x-33)+4 (5 N-3 x+18)\right)}\delta u'+$}\\
&+\frac{16 x \left(2 d^3-d^2 (3 x+8)+d (4 N-66)+20 N+12 (x+6)\right)}{(d+2) \left(2 d^3-d^2 (3 x+8)+2 d
   (2 N-6 x-33)+4 (5 N-3 x+18)\right)}\delta u''=0\\
\end{split}\end{equation}
and
\begin{equation}\begin{split}
&(d-\theta -2) \delta f+\\
&\resizebox{1.05\hsize}{!}{$+\frac{\left(-4 N \left(d^3 (x-4)+d^2 (11 x+16)+4 d
   (5 x+33)+4 (x-36)\right)-(d+2) x \left(2 d^4-3 d^3 (x+4)-2 d^2 (3 x+25)+12 d (x+17)+24 (x-10)\right)+32 (d+5) N^2\right)
   }{(d+2) \left(2 d^3-d^2 (3 x+8)+2 d (2 N-6 x-33)+4 (5 N-3 x+18)\right)}\delta f'+$}\\
&+\frac{16 x \left(2 d^3-d^2 (3
   x+8)+d (4 N-66)+20 N+12 (x+6)\right)}{(d+2) \left(2 d^3-d^2 (3 x+8)+2 d (2 N-6 x-33)+4 (5 N-3
   x+18)\right)}\delta f''-\\
&-\frac{16 (d+5) \left(N \left(2 d^3-d^2 (3 x+8)-6 d (2 x+11)-12 (x-6)\right)+4 (d+5) N^2+12 (d+2) x\right)}{3 (d-1) (d+2)^2 \left(2 d^3-d^2 (3 x+8)+2 d (2 N-6 x-33)+4 (5 N-3 x+18)\right)}\delta u'-\\
&-\frac{32 (d+5) x \left(2 d^3-d^2 (3 x+8)+d (4 N-66)+20 N+12 (x+6)\right)}{3 (d-1)
   (d+2)^2 \left(2 d^3-d^2 (3 x+8)+2 d (2 N-6 x-33)+4 (5 N-3 x+18)\right)}\delta u''=0
\end{split}\end{equation}
the equation of $\delta u$ is independent of $\delta f$, accordingly the critical exponents are determined only by the potential. Using the Frobenius method with ansats $\delta u(x)=\sum_{i=0}^{\infty}u_i(x-x_0)^i$ shows that all coefficients $u_i$ contain a denominator given by $d (d (2 d-3 x_0-8)+4 N-66)+4 (5 N+3 (x_0+6))$. If this denominator goes to zero the solution has a discontinuity. With $N<N_c$ of eq.(\ref{NCs1}) the discontinuity is located at a negative number $x_0<0$ but for $N\ge N_c$ the zero of the denominator moves to positive numbers. This implies that there are smooth solutions only for $N<N_c$. Smooth solutions with $N< N_c$ can be found by the shooting technique described in Appendix \ref{appendix4}. Tables \ref{critS23} and \ref{critS24} show the critical exponents in $d=3$ and $d=4$. In the case $N=1$ the results for the relevant directions are the same of \cite{Percacci:2015wwa}. 

\begin{table}[t]
  \centering
  \begin{minipage}{0.48\textwidth}
    \centering
    \caption{Critical exponents related to the scaling solution eq.(\ref{scal1}) in $d=3$.}
    \label{critS23}
    $\begin{array}{|c|ccccc|}
    \hline
     N & \theta _1 & \theta _2 & \theta _3 & \theta _4 & \theta _{-1} \\
    \hline
     1 & 3.000 & 1.790 & 1.000 & 0.497 & -0.798 \\
     2 & 3.000 & 1.880 & 1.000 & 0.701 & -0.494 \\
     3 & 3.000 & 1.950 & 1.000 & 0.870 & -0.224 \\
     4 & 3.000 & 1.990 & 1.000 & 0.975 & -0.040 \\
    \hline
    \end{array}$
  \end{minipage}%
  \hfill
  \begin{minipage}{0.48\textwidth}
    \centering
    \caption{Critical exponents related to the scaling solution eq.(\ref{scal1}) in $d=4$.}
    \label{critS24}
    $\begin{array}{|c|cccc|}
    \hline
     N & \theta _1 & \theta _2 & \theta _3 & \theta_{-1} \\
    \hline
     1 & 4.000 & 2.000 & 1.770 & -0.643 \\
     2 & 4.000 & 2.000 & 1.830 & -0.470 \\
     3 & 4.000 & 2.000 & 1.880 & -0.302 \\
     4 & 4.000 & 2.000 & 1.940 & -0.149 \\
     5 & 4.000 & 2.000 & 1.990 & -0.028 \\
    \hline
    \end{array}$
  \end{minipage}
\end{table}

The perturbations cannot be computed analytically, except in the asymptotic regimes. The asymptotics at $x\to x_0$ are the standard Frobenius expansions. In the asymptotic regime $x\to\infty$ the solutions are given by
\begin{equation}\begin{split}
\label{pertinfS2}
&\delta u(x\to\infty)=x^{\frac{d-\theta}{d-2}} \left(c_1+\sum _{n=1}^\infty v_{-n} x^{-n}\right)\\ 
&\delta f(x\to\infty)=x^{1-\frac{\theta }{d-2}} \left(c_2+f_{4-d} x^{4-d}+\sum _{n=1}^{\infty} f_{-n} x^{-n}+\log (x) \sum _{n=0}^\infty g_{-n}
   x^{-n}\right)
\end{split}\end{equation}
where $c_1$ and $c_2$ are two free coefficients. These coefficients are determined numerically by the shooting. The coefficients $v_i$, $f_i$ and $g_i$ are determined by the Frobenius method. In $d=4$ the first coefficients are given by
\begin{equation}\begin{split}
&v_{-1}=\frac{1}{9} c_1 (\theta -4) (3 N-\theta )\\
&v_{-2}=\frac{1}{162} c_1 (\theta -4) \left(\theta ^3-36 \theta +9 (\theta -2) N^2-6 (\theta -1)^2 N+96\right)
\end{split}\end{equation}
and 
\begin{equation}\begin{split}
&f_{-1}=-\frac{2}{243} c_1 (\theta -4) \left(2 (\theta -8) \theta +9 N^2-3 (2 \theta +3) N+48\right)-\frac{1}{9} c_2 (\theta -2)
   (\theta -3 N+2)\\
&f_{-2}=\frac{1}{162} c_2 (\theta -2) \left(\theta  (\theta  (\theta +6)-24)+9 \theta  N^2-6 (\theta +1)^2
   N+32\right)+\\
   &\resizebox{1.1\hsize}{!}{$+\frac{c_1}{2187} (\theta -4) \left(4 \theta  (\theta  ((\theta -5) \theta -18)+24)-54 (\theta -1) N^3+9 \left(6
   \theta ^2+3 \theta -4\right) N^2-3 (\theta  (4 \theta  (2 \theta -5)+3)+160) N+768\right)$}\\
&g_{0}=\frac{1}{27} c_1 (\theta -4) (3 N-\theta ),\quad \quad g_{-1}=\frac{1}{243} c_1 (\theta -4) (\theta -2) (-\theta +3 N-2) (3 N-\theta )\\
&g_{-2}=\frac{c_1}{4374}(\theta -4) (\theta -2) (3 N-\theta ) \left(\theta  (\theta  (\theta +6)-24)+9 \theta  N^2-6 (\theta +1)^2
   N+32\right)
\end{split}\end{equation}

\section{The large \texorpdfstring{$N$}{N} limit}\label{secNinf}
\noindent The large $N$ limit of a quantum field theory is of particular interest because the phase structure of the theory can be studied analytically (look at \cite{Moshe:2003xn} for a review and other applications). In this section we study the limit $N\to\infty$ of eqs.(\ref{betau}) and (\ref{betaf}).

\subsection{The scaling solutions of \texorpdfstring{$N=\infty$}{N}}
In the limit $N\to\infty$ the model can be solved exactly, since it reduces to the so-called spherical model \cite{PhysRevA.8.401,Comellas:1997tf}. It is convenient to rescale $x$, $u_*$ and $f_*$ by $4N$ and obtain in the large $N$ limit the simplified flow equations
\begin{equation}\begin{split}
&\dot{u}=-d u_*+(d-2) x u'_*+\left(1+\frac{u'_*}{m}\right)^{\frac{d}{2}-m}\\
&\dot{f}=(2-d) f_*+(d-2) x f'_*+\left(\frac{d}{2m}-1\right)\left(\frac{1}{6}+f_\ast^\prime\right)\left(1+\frac{u_\ast^\prime}{m}\right)^{\frac{d}{2}-m-1} \, .
\end{split}\end{equation}
In this limit gravity does not affect the fixed-point potential, which coincides with the flat spacetime result \cite{Mazza:2001bp}. Furthermore, there is no difference between the C- and B- type cutoff, since the $\epsilon$ dependence disappears.

The corresponding fixed point equations admit a solution with constant $u_*$ and $f_*$, which coincide with the limit $N\to\infty$ of eq.(\ref{GFP}) rescaled by $4N$. 

The fixed point equations also admit a non-trivial solution, which can be obtained easily moving to study the fixed point equations for $u'_*$ and $f_*$:
\begin{equation}\begin{split}
\label{solanN}
&x=c_1(u_\ast^\prime)^{\frac{d}{2}-1}-\frac{1}{d-2}\left(\frac{d}{2m}-1\right) {}_2F_1{\left(1-\frac{d}{2},-\frac{d}{2}+m+1;2-\frac{d}{2};-\frac{u_\ast^\prime}{m}\right)}\\
&f_*\left(u_\ast^\prime\right)=-\frac{x}{6}+\left(\frac{c_1}{6}+c_2\right)(u_\ast^\prime)^{\frac{d}{2}-1}
\end{split}\end{equation}
The solution of $u_*$ is given in an implicit form in terms of its derivative and the solution of $f_*$ is expressed in terms of this derivative. $c_1$ and $c_2$ are two free parameters. These solutions are defined for $d\ne4$ and for $d=4$ only the gaussian fixed-point remains. However, from a more careful inspection a non-trivial solution globally defined and with a potential bounded from below can exist only for $d<4$ (see for example~\cite{Percacci:2014tfa}).

To extract the scaling solutions from eq. (\ref{solanN}) we require the smoothness for all values of $x$. The smoothness at $x=0$ requires $u_*'(0)$ and $f_*(0)$ to be set to some value $v_0$ and $f_0$, this fixes $c_1$ and $c_2$. The smoothness at some generic $x=x_0$ can be studied by Taylor expansion. This shows that $c_1$ and $c_2$ have to be set to zero to get a smooth solution. At the end there is only one scaling solution, where $f_0$ is zero and $v_0$ is given by the solution of 
\begin{equation}
\label{solv0Ninf}
{}_2 F_1{\left(1-\frac{d}{2},-\frac{d}{2}+m+1;2-\frac{d}{2};-\frac{v_0}{m}\right)}=0
\end{equation}
for example, with $d=3$ and $m=d/2+1$ we get $v_0\sim-0.97$. The scaling solutions are given by
\begin{equation}\begin{split}
\label{scalNinf}
&x=-\frac{1}{d-2}\left(\frac{d}{2m}-1\right) {}_2F_1{\left(1-\frac{d}{2},-\frac{d}{2}+m+1;2-\frac{d}{2};-\frac{u_\ast^\prime}{m}\right)}\\
&f_*=-\frac{x}{6}
\end{split}\end{equation}
the solution for $u_*'(x)$ is the Heisenberg fixed-point \cite{PhysRevA.8.401,Comellas:1997tf}. In this limit $f_*$ is always negative, this is also what was found in the analysis based on the 1PI effective average action in \cite{Labus:2015ska}, with the difference that the function $f_*$ was a non trivial function, having a linear behavior only at asymptotically large values of $x$. Moreover, in that analysis $f_*(0)$ was found to have a negative value, whereas here in the Wilsonian proper time framework we find $f_*(0)=0$.

\subsection{The critical exponents of \texorpdfstring{$N=\infty$}{N}}
The linearized equations around the scaling solutions are given by
\begin{equation}\begin{split}
&(d-\theta )\delta u+ \delta u' \left(\left(1-\frac{d}{2m}\right)\left(1+\frac{u_\ast^\prime}{m}\right)^{\frac{d}{2}-m-1}- (d-2)x\right)=0\\
&(d-\theta -2) \delta f+ \delta f' \left(\left(1-\frac{d}{2m}\right)\left(1+\frac{u_\ast^\prime}{m}\right)^{\frac{d}{2}-m-1}-(d-2)x\right)=0
\end{split}\end{equation}
this first order system can be solved analytically, in particular the two equations are the same if in the second $\theta+2\to\theta$.

The critical exponents and their respective eigenfunctions for the gaussian-fixed point are given by the $N\to\infty$ limit of eq.(\ref{CritExpGFP}) and 
eqs.(\ref{ein1GFP}), (\ref{ein2GFP}), after having multiplied by $4N$.

To get the critical exponents for the non-trivial scaling solution, we require that the two equations have no singularity for all values of $x$. This means that the terms involving the first derivative should be different from zero for all values of $x$. Using $\delta u=\sum _{j=n}^{\infty} a_j (x-x_0)^j$ to study the behavior around a generic $x_0$, the previous condition is met only if 
\begin{equation}
\label{critNinf}
\theta=d-2j, \quad\quad j=0,1,...
\end{equation}
which is the well-know spectrum for the spherical model. The perturbations can be found analytically in terms of $u_*'$
\begin{equation}
\delta u\left(u_\ast^\prime\right)=b_1 (u_\ast^\prime)^\frac{d-\theta}{2},\quad \quad \delta f\left(u_\ast^\prime\right)=b_2(u_\ast^\prime)^\frac{d-\theta-2}{2}
\end{equation}
where $b_1$ and $b_2$ are two free parameters. 

\section{The gravitationally dressed Wilson-Fisher fixed-point}\label{secGWF}
\noindent In this section we discuss the non-trivial scaling solutions in $d=3$. A search of a gravitationally-dressed Wilson-Fisher fixed point has also been done in \cite{Percacci:2015wwa}, \cite{Borchardt:2015rxa} with $N=1$ and in \cite{Labus:2015ska} with $N=2$ based on the RG flow of the 1PI effective average action.

\subsection{Numerical techniques: shooting vs pseudospectral method}\label{secSvsPS}
The most widely used numerical technique for finding non-trivial scaling solutions is the shooting method \cite{Morris:1994ie, Bonanno:2000yp, Litim:2010tt, Mazza:2001bp, Narain:2009fy, Percacci:2015wwa}. Also, recently in \cite{Borchardt:2015rxa} a technique based on the pseudospectral method has been used. With the propertime flow equation we use both techniques to have a double check of the results. For better convergence and to have a less numerical error, we exploit an "improved version" of the shooting and pseudospectral method. We describe the numerical techniques in the appendix \ref{appendix4}.

The shooting technique and the pseudospectral method have different sources of numerical error. However, both methods share a common source: the truncation of a power series —around $x\to\infty$ in the shooting method, and in the Chebyshev expansion in the pseudospectral method. This truncation affects the numerical values of the solutions. Additionally, in the pseudospectral method, it can also lead to the presence of spurious solutions, which must be carefully distinguished from the physical ones. For these reasons, we consider the shooting method as our primary technique for numerical computations, using the pseudospectral method mainly as a cross-check for the shooting results.

To distinguish physical from spurious solutions in the pseudospectral method, we study the results for different values of the truncation $p$ of Chebyshev expansion. In particular, we use the relative difference $\delta u_*^{PS}(p,x)=|\frac{u_*(p+1,x)-u_*(p,x)}{u_*(p,x)}|$ as an error estimate test at different values of $x$. For a true solution $\delta u_*^{PS}(p\to\infty,x)\to0$.

As explained in appendix \ref{appendix4}, due to our implementation of the shooting technique, the truncation does not affect the solution in a critical way and the only potentially problematic source of error may arise from systematic errors during the numerical integration. These errors could lead to spurious numerical convergence in the solutions of eq.(\ref{sysnum}). To test whether the solutions found by the shooting method correspond to genuine nontrivial fixed-point solutions, we fit them to the pseudospectral ansatz, eq.(\ref{PSans}), and attempt to recover them using the pseudospectral method for different values of the truncation order $p$. We use the relative difference $\delta u_*^{SPS}(p,x)=|\frac{u_*^{shoot}(x)-u_*^{PS}(p,x)}{u_*^{shoot}(x)}|$ to test the results. For illustration, we focus on $u_*$, but similar conclusions hold for $f_*$. 

\begin{figure}[t]
     \centering
     \subfigure[]{
         \centering
         \includegraphics[width=0.45\textwidth]{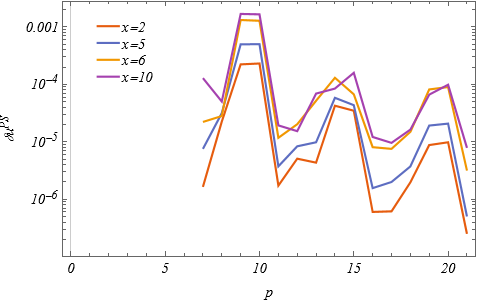}
         \label{diffPS}
     }
    \hspace{0.8em}
     \subfigure[]{
         \centering
         \includegraphics[width=0.45\textwidth]{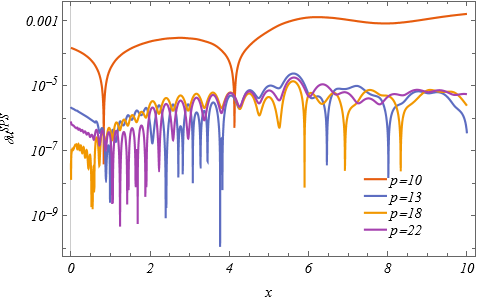}
         \label{diffSPS}
     }
\caption{Log plot of the relative difference $\delta u_*^{PS}(p)$ in the C-type cutoff for different values of $x$ in (a) and $\delta u_*^{SPS}(x)$ for different values of $p$ in (b).}
\end{figure}

Fig.~\ref{diffPS} shows the log plot of the relative difference $\delta u_*^{PS}(p,x)$ for the scaling solution of C-type cutoff ($\epsilon=0$) with $N=1$ and $m=d/2+1$ for different values of $x$. The relative difference is around $10^{-1}$ for $p<10$ and decreases as $p$ increases. In particular, the larger $x$ is, the bigger the relative difference tends to be. For $p>20$ the relative difference remains always below $10^{-5}$ for all tested values of $x$. 

Fig.~\ref{diffSPS} shows the log plot of relative difference $\delta u_*^{SPS}(p,x)$ for different values of $p$. For all values of $x$ the relative difference $\delta u_*^{SPS}(p,x)$ decreases as $p$ increases. With $p>20$ the value is always below $10^{-6}$ and there is no substantial difference between the solution of pseudospectral and shooting method. By further increasing the truncation both $\delta u_*^{PS}(p,x)$ and $\delta u_*^{SPS}(p,x)$ decrease more and more, which confirms that the solution of the shooting method is a genuine fixed-point solution. The same situation occurs for other values of $m$ and $N$.

\begin{figure}[t]
\vspace{0.3cm}
     \centering
    \subfigure[]{
        \includegraphics[width=0.45\textwidth]{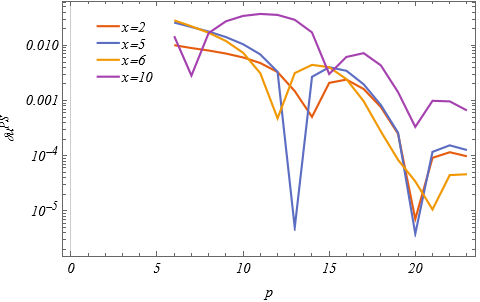}
        \label{diffPSeps1}
    }
    \hspace{0.8em}
     \subfigure[]{
         \includegraphics[width=0.45\textwidth]{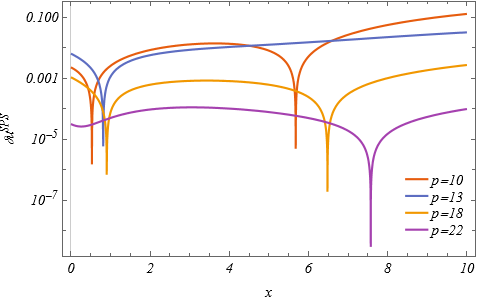}
         \label{diffSPSeps1}
     }
\caption{Log plot of the relative difference $\delta u_*^{PS}(p)$ in the B-type cutoff for different values of $x$ in (a) and $\delta u_*^{SPS}(x)$ for different values of $p$ in (b).}
\end{figure}

In the B-type cutoff ($\epsilon=1$) the differential equations are much more complicated to solve. The numerical technique follows the same line as the C-type cutoff. Figs. \ref{diffPSeps1} and \ref{diffSPSeps1} show the relative differences. Here the error is slightly larger, but again we find that the pseudospectral method and the shooting results agree when $p>20$. 

\subsection{Scaling solutions for scheme C}
\begin{figure}[p]
     \centering
     \subfigure[]{
         \centering
         \includegraphics[width=0.45\textwidth]{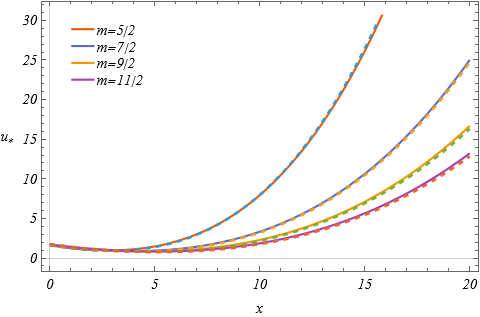}
         \label{plotu}
     }
    \hspace{0.8em}
     \subfigure[]{
         \centering
         \includegraphics[width=0.45\textwidth]{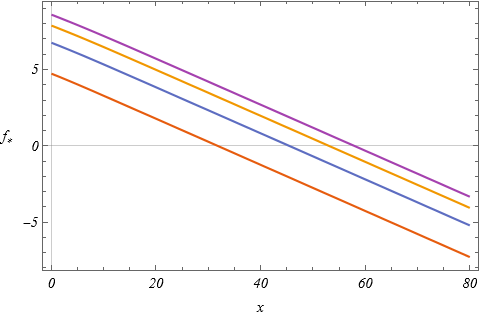}
         \label{plotf}
     }
\caption{Scaling solutions for $u_*$ and $f_*$ for $N=1$ and different values of $m$ in the C-type cutoff. The dashed lines in (a) are the Wilson-Fisher scaling solutions of C-type cutoff obtained from eq.(\ref{betau}) when $f_*=0$.}

\vspace{0.5cm}

     \centering
     \subfigure[]{
         \centering
         \includegraphics[width=0.45\textwidth]{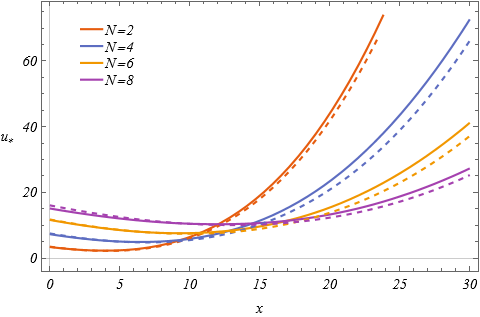}
         \label{plotNu}
     }
    \hspace{0.8em}
     \subfigure[]{
         \centering
         \includegraphics[width=0.45\textwidth]{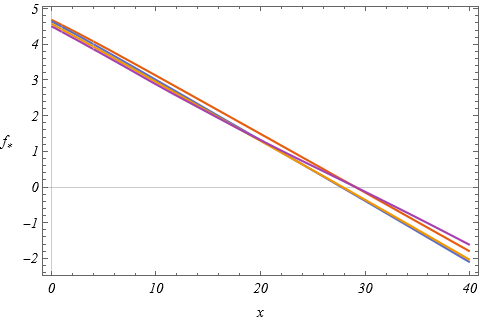}
         \label{plotNf}
     }
\caption{Scaling solutions for $u_*$ and $f_*$ for different values of $N$ and $m=d/2+1$ in the C-type cutoff. The dashed lines in (a) are the Wilson-Fisher scaling solutions of C-type cutoff obtained from eq.(\ref{betau}) when $f_*=0$.}

\vspace{0.5cm}

     \centering
     \subfigure[]{
         \centering
         \includegraphics[width=0.45\textwidth]{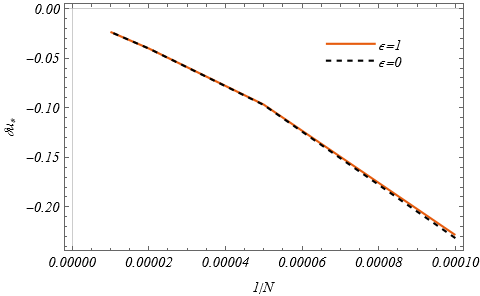}
         \label{plotulargeN}
     }
    \hspace{0.8em}
     \subfigure[]{
         \centering
         \includegraphics[width=0.45\textwidth]{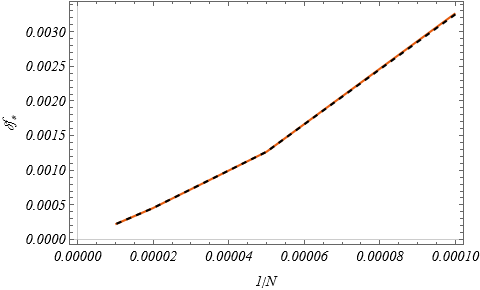}
         \label{plotflargeN}
     }
\caption{Plot of the difference $\delta u_N=u_*(x,N\to\infty)-u_*(x,N)$ in the left and $\delta f_N=f_*(x,N\to\infty)-f_*(x,N)$ in the right as function of $1/N$ at $x=3$. The dashed lines are the results in the C-scheme, the full lines the results in the B-scheme.}
\end{figure}

Figs.~\ref{plotu} and \ref{plotf} show the scaling solutions of the C-type cutoff for $N=1$ and different values of $m$. The scaling solutions for $f_*$ in fig.~\ref{plotf} follow approximately straight lines of negative slope, accordingly at some $\bar{x}$ they cross the $x$-axis and become negative. These solutions can be approximated by $f_*(x)\sim \frac{1}{16\pi g_*^{PG}}+f_\infty x$.

The dashed lines in fig.~\ref{plotu} are the pure Wilson-Fisher scaling solutions, obtained from eq.(\ref{betau}) when $f_*=0$. The gravitationally dressed WF $u_*$ are very similar to them. 

In the WF case $u'(x=0)$ is related to the critical mass $m_c^2$. This value is negative. In presence of gravity in the C-type cutoff $m_c^2$ is still negative.

Fig.~\ref{plotNu} and \ref{plotNf} show the scaling solutions with $m=d/2+1$ for $N>2$. The results  follow the same trend of $N=1$. The gravitational potential and the pure potential are again very similar for all values of $N$. As $N$ increases, $f_*$ shifts to smaller values. This has to be expected since in the limit $N\to\infty$ $f_*(x)\to-2N x/3$. 

Using $m=d/2+1$, for $N>9$ we do not find real scaling solutions until $N\sim8184$ is reached. The numerical scaling solutions with $N\gtrsim8184$ tend to the scaling of $N\to\infty$ in eq. (\ref{scalNinf}). To quantify the differences, we plot $\delta u_N=u_*(x,N\to\infty)-u_*(x,N)$ and $\delta f_N=f_*(x,N\to\infty)-f_*(x,N)$ as functions of $1/N$ for different values of $x$. The dashed lines in figs. \ref{plotulargeN} and \ref{plotflargeN} show the results at $x=3$, $\delta u_N$ and $\delta f_N$ follow approximately straight lines, so the corrections to eq.(\ref{scalNinf}) go as $1/N$. 

\subsection{Scaling solutions for the scheme B}
\begin{figure}[t!]
     \centering
     \subfigure[]{
         \centering
         \includegraphics[width=0.45\textwidth]{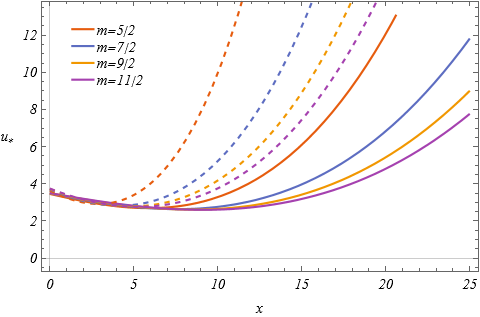}
         
         \label{plotueps1}
     }
    \hspace{0.8em}
     \subfigure[]{
         \centering
         \includegraphics[width=0.45\textwidth]{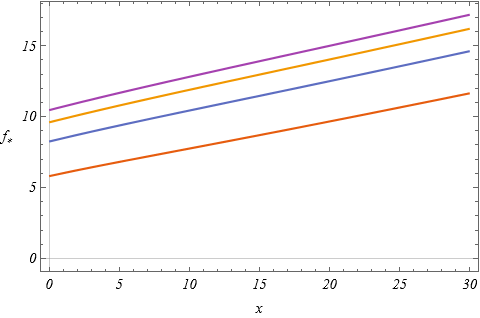}
         
         \label{plotfeps1}
     }
\caption{Scaling solutions for $u_*$ and $f_*$ for $N=1$ and different values of $m$ in the B-type cutoff. The dashed lines in (a) are the Wilson-Fisher scaling solutions of B-type cutoff obtained from eq.(\ref{betau}) when $f_*=0$.}

\vspace{1cm}

     \centering
     \subfigure[]{
         \centering
         \includegraphics[width=0.45\textwidth]{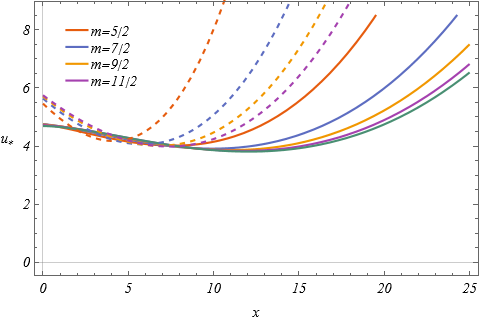}
         
         \label{plotNueps1}
     }
    \hspace{0.8em}
     \subfigure[]{
         \centering
         \includegraphics[width=0.45\textwidth]{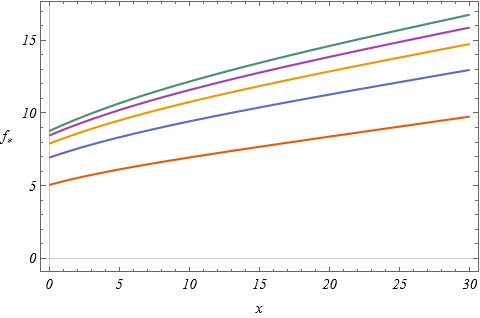}
         
         \label{plotNfeps1}
     }
\caption{Scaling solutions for $u_*$ and $f_*$ for $N=2$ and different values of $m$ in the B-type cutoff. The dashed lines in (a) are the Wilson-Fisher scaling solutions of B-type cutoff obtained from eq.(\ref{betau}) when $f_*=0$.} 
\end{figure}

Figs. \ref{plotueps1}, \ref{plotfeps1} and \ref{plotNueps1}, \ref{plotNfeps1} show the scaling solutions of the B-type cutoff with $N=1$ and $N=2$ for different values of $m$. 

In figs. \ref{plotueps1} and \ref{plotNueps1} the plots show $u_*$ and the dashed line are again the pure Wilson-Fisher case in absence of gravitational interactions. In contrast to the C-type cutoff, in the B-type gravity the minimum of $u_*$ shifts to a larger value and the similarity with the WF case is lost.

In figs. \ref{plotfeps1} and \ref{plotNfeps1} the plots show $f_*$. The solutions $f_*$ show a positive slope. This same trend is observed in the scaling solution for the 1PI effective average action which satisfy the Wetterich-Morris equation, as was shown in \cite{Borchardt:2015rxa}. In particular, the solution with $m=d/2+1$ is the one that most resembles it.

For both values of $N$ the critical mass $u'(0)=m_c^2$ takes a negative value.

Using $m=d/2+1$, for $N>2$ we do not find real scaling solutions until $N\sim10000$ is reached. The numerical scaling solutions with $N\gtrsim10000$ tend to the scaling of $N\to\infty$ in eq.(\ref{scalNinf}). The full lines in figs. \ref{plotulargeN} and \ref{plotflargeN} show the result for the differences $\delta u_N$ and $\delta f_N$. As in the C-scheme the corrections go as $1/N$, in particular the difference between the two schemes is very small and decreases more and more as $N$ increases.

\section{stability analysis for the gravitationally-dressed Wilson-Fisher fixed-point}\label{seccritWF}
\noindent Linearizing the flow equations around the gravitational WF fixed-point we get the critical properties. The linearized equations are given in the Appendix \ref{appendix3}. To solve these equations, we exploit again the shooting and use the pseudospectral method as a check. However, in addition, to compute the critical exponents, we exploit the polynomial truncation around the minimum of the scaling $u_*$. This technique leads to very accurate critical exponents and acts as an independent check of the results. The technique is described in Appendix \ref{appendixPolTr}. 

\subsection{Critical exponents for the scheme C}
In the C-type cutoff the similarity of the gravitational result with the classical WF case reflects also in the flow around the fixed-point. In particular, as in the classical WF case, there is only one non-trivial relevant direction $\theta_1$ where one sets $\nu=1/\theta_1$. The first irrelevant direction is labeled by $\omega$.

\begin{figure}[t]
     \centering
     \subfigure[]{
         \centering
         \includegraphics[width=0.45\textwidth]{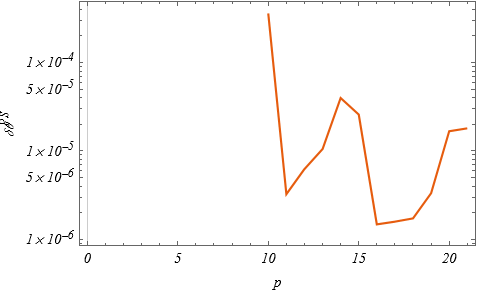}
         
         \label{difflambdapos}
     }
    \hspace{0.8em}
     \subfigure[]{
         \centering
         \includegraphics[width=0.45\textwidth]{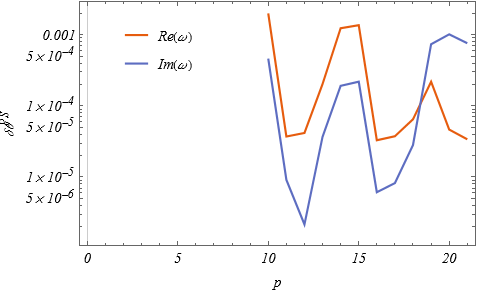}
         
         \label{difflambdaneg}
     }
\caption{Log plot of the relative difference $\delta\theta^{PS}$ in the C-type cutoff related to $\theta_1=1/\nu$ in (a) and $\omega$ in (b), both for the imaginary and real part of $\omega$.}

\vspace{1cm}

     \centering
     \subfigure[]{
         \centering
         \includegraphics[width=0.45\textwidth]{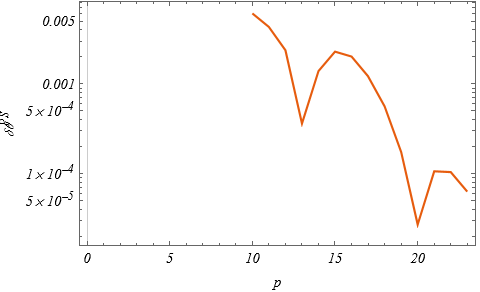}
         
         \label{difflambdaposeps1}
     }
    \hspace{0.8em}
     \subfigure[]{
         \centering
         \includegraphics[width=0.45\textwidth]{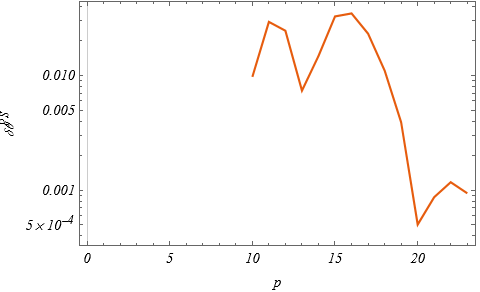}
         
         \label{difflambdanegeps1}
     }
\caption{Log plot of the relative difference $\delta\theta^{PS}$ in the B-type cutoff related to $\theta_1$ in (a) and $\omega$ in (b).}
\end{figure}

The numerical approach we employ to get and check the eigenfunctions follows the same logic as in Subsection \ref{secSvsPS}. In particular, now we also consider the relative difference $\delta\theta^{PS}=|\frac{\theta^{PS}(p+1)-\theta^{PS}(p)}{\theta^{PS}(p)}|$ to test the precision of the critical exponents. Fig. \ref{difflambdapos} and \ref{difflambdaneg} show the log plot of $\delta\theta^{PS}$ for $\theta_1=1/\nu$ and  $\omega$. The relative difference decreases as $p$ increases, for $p>20$ this is below $10^{-5}$ and $10^{-4}$ for the positive and negative critical exponent, respectively. Similar results are also obtained for $N>1$ and other values of $m$. 

The trivial relevant directions are $\theta=3$ and $\theta=1$. Tables \ref{tablenu}, \ref{tableomega} and \ref{CritExpN>4C} show the critical exponents $\nu$ and $\omega$ for different numbers of $N$ and $m$ and the comparison with the Wilson-Fisher case. As $m$ increases, the values tend to those of $m\to\infty$. The gravitational $\nu$s are very close to the WF case. 
In contrast, $\omega$ shows big deviations, for $N=1$ and $N=2$ there is an imaginary part. The imaginary part disappears for $N>2$. 

In \cite{Percacci:2015wwa} with the unimproved Wetterich equation for $N=1$ similar results were found, where $\omega_{grav}$ also has an imaginary part. Our results differ from those of \cite{Percacci:2015wwa} for less than one percent. 

The perturbations cannot be computed analytically except in the asymptotic regimes. Around a point $x=x_0$ the solutions are the standard Frobenius expansions. Around $x\to\infty$ the solutions are the same form of eq.(\ref{pertinfS2}) without the logarithmic term. In Appendix \ref{applinasin} we give the explicit expressions for the series around $x=0$ and $x\to\infty$.

\begin{table}[p]
\caption{Comparison between the numerical results in the C-type cutoff for $\nu$ for different values of $m$ and $N$.}
\label{tablenu}
$\begin{array}{|c|cc @{\hspace{1cm}} cc @{\hspace{1cm}} cc @{\hspace{1cm}} cc|}
\hline
  & N=1 &  & N=2 &  & N=3 &  & N=4 &  \\
\hline
m & \nu _{\text{WF}} & \nu _{\text{grav}} & \nu _{\text{WF}} & \nu _{\text{grav}} & \nu _{\text{WF}} & \nu
   _{\text{grav}} & \nu _{\text{WF}} & \nu _{\text{grav}} \\
5/2 & 0.650 & 0.639 & 0.708 & 0.692 & 0.761 & 0.745 & 0.804 & 0.794 \\
 7/2 & 0.640 & 0.632 & 0.693 & 0.682 & 0.743 & 0.734 & 0.785 & 0.784 \\
9/2 & 0.636 & 0.629 & 0.687 & 0.677 & 0.734 & 0.728 & 0.776 & 0.780 \\
 11/2 & 0.634 & 0.628 & 0.683 & 0.674 & 0.729 & 0.725 & 0.770 & 0.777 \\
 13/3 & 0.633 & 0.627 & 0.680 & 0.672 & 0.726 & 0.723 & 0.767 & 0.775 \\
 15/2 & 0.632 & 0.626 & 0.679 & 0.671 & 0.723 & 0.722 & 0.764 & 0.773 \\
 17/2 & 0.631 & 0.625 & 0.677 & 0.670 & 0.722 & 0.720 & 0.762 & 0.773 \\
 19/2 & 0.630 & 0.625 & 0.676 & 0.670 & 0.720 & 0.719 & 0.760 & 0.774 \\
 \hline
\end{array}$

\vspace{0.5cm}

\caption{Comparison between the numerical results in the C-type cutoff for $\omega$ for different values of $m$ and $N$.}
\label{tableomega}
$\begin{array}{|c|cc @{\hspace{1cm}} cc @{\hspace{1cm}} cc @{\hspace{1cm}} cc|}
\hline
  & N=1 &  & N=2 &  & N=3 &  & N=4 &  \\
\hline
 m & \omega _{\text{WF}} & \omega _{\text{grav}} & \omega _{\text{WF}} & \omega _{\text{grav}} & \omega _{\text{WF}} &
   \omega _{\text{grav}} & \omega _{\text{WF}} & \omega _{\text{grav}} \\
 5/2 & 0.656 & 0.589\pm 0.139 i & 0.671 & 0.627\pm 0.102 i & 0.700 & 0.648 & 0.734 & 0.636 \\
 7/2 & 0.688 & 0.595\pm 0.136 i & 0.689 & 0.623\pm 0.100 i & 0.702 & 0.615 & 0.723 & 0.605 \\
 9/2 & 0.705 & 0.599\pm 0.134 i & 0.701 & 0.622\pm 0.098 i & 0.706 & 0.601 & 0.721 & 0.591 \\
 11/2 & 0.716 & 0.602\pm 0.132 i & 0.708 & 0.623\pm 0.096 i & 0.710 & 0.594 & 0.721 & 0.582 \\
 13/2 & 0.724 & 0.604\pm 0.130 i & 0.714 & 0.623\pm 0.094 i & 0.713 & 0.588 & 0.721 & 0.574 \\
 15/2 & 0.729 & 0.605\pm 0.129 i & 0.718 & 0.623\pm 0.093 i & 0.716 & 0.585 & 0.725 & 0.573 \\
 17/2 & 0.733 & 0.606\pm 0.128 i & 0.721 & 0.623\pm 0.092 i & 0.718 & 0.583 & 0.722 & 0.564 \\
 19/2 & 0.736 & 0.607\pm 0.127 i & 0.724 & 0.624\pm 0.091 i & 0.719 & 0.580 & 0.723 & 0.579 \\
 \hline
\end{array}$

\vspace{0.5cm}
\caption{Critical exponents with $m=d/2+1$ for $N=5,6,7,8,9$ in the C-type cutoff and comparison with WF case.}
\label{CritExpN>4C}
$\begin{array}{|cc @{\hspace{1cm}} cc @{\hspace{1cm}} cc @{\hspace{1cm}} cc @{\hspace{1cm}} cc|}
\hline
 N=5 &  & N=6 &  & N=7 &  & N=8 &  & N=9 &  \\
\hline
 \nu _{\text{WF}} & \nu _{\text{grav}} & \nu _{\text{WF}} & \nu _{\text{grav}} & \nu _{\text{WF}} & \nu _{\text{grav}} & \nu
   _{\text{WF}} & \nu _{\text{grav}} & \nu _{\text{WF}} & \nu _{\text{grav}} \\
 0.838 & 0.847 & 0.863 & 0.887 & 0.882 & 0.827 & 0.897 & 0.822 & 0.909 & 0.912\\[5pt]

\omega _{\text{WF}} & \omega _{\text{grav}} & \omega _{\text{WF}} & \omega _{\text{grav}} & \omega _{\text{WF}} & \omega
   _{\text{grav}} & \omega _{\text{WF}} & \omega _{\text{grav}} & \omega _{\text{WF}} & \omega _{\text{grav}} \\
 0.767 & 0.793 & 0.796 & 0.626 & 0.820 & 0.645 & 0.841 & 0.771 & 0.859 & 0.939 \\
\hline
\end{array}$
\end{table}

\subsection{Critical exponents for the scheme B}
In the B-type cutoff the large difference between the classical WF potential and the gravitational WF case is also reflected in the flow around the fixed point.

To test the precision of the results we use again the relative difference $\delta\theta^{PS}$. Figs. \ref{difflambdaposeps1} and \ref{difflambdanegeps1} show $\delta\theta^{PS}(p)$ for the first positive and negative critical exponents as a log plot. In contrast to the C-type cutoff here, the error is slightly larger, for $p>20$ the relative difference is $\sim10^{-4}$ and $10^{-3}$ for the positive and negative critical exponents, respectively.

Table \ref{tableeps1N1} shows the numerical results for the critical exponents for $N=1$, $2$ and different values of $m$. Apart from the trivial relevant direction $\theta=3$, with $N=1$ we find three nontrivial relevant directions. The critical exponents in this study of the Wilsonian action with proper time regulator are found to be slightly larger than the ones obtained in \cite{Borchardt:2015rxa} for the 1PI effective average action. 

\begin{table}[t]
\caption{Numerical results for the critical exponents in the B-type cutoff in the case $N=1$ and $N=2$ for different values of $m$.}
\label{tableeps1N1}
$\begin{array}{|c|cc @{\hspace{1cm}} cc @{\hspace{1cm}} cc @{\hspace{1cm}} cc|}
\hline
  & \nu _{\text{grav}} &  & \theta _2 &  & \theta _3 &  & \omega _{\text{grav}} &  \\
\hline
 m & N=1 & N=2 & N=1 & N=2 & N=1 & N=2 & N=1 & N=2 \\
  5/2 & 0.467 & 0.438 & 1.140 & \text{} & 1.090 & 1.140\pm 0.164 i & 0.290 & 0.156 \\
 7/2 & 0.485 & 0.449 & 1.220 & \text{} & 1.000 & 1.110\pm 0.129 i & 0.307 & 0.128 \\
 9/2 & 0.491 & 0.452 & 1.230 & \text{} & 0.978 & 1.100\pm 0.105 i & 0.312 & 0.099 \\
 11/2 & 0.495 & 0.453 & 1.240 & \text{} & 0.843 & 1.100\pm 0.074 i & 0.315 & 0.072 \\
 13/2 & 0.497 & 0.453 & 1.240 & 1.150 & 0.829 & 1.070 & 0.317 & 0.035 \\
 15/2 & 0.499 & 0.453 & 1.240 & 1.200 & 0.819 & 1.020 & 0.318 & 0.009 \\
 17/2 & 0.500 & 0.454 & 1.250 & 1.210 & 0.812 & 1.010 & 0.319 & 0.004 \\
 19/2 & 0.501 & 0.455 & 1.250 & 1.220 & 0.807 & 1.000 & 0.319 & 0.005 \\
 \hline
\end{array}$
\end{table}

In contrast to the C-type cutoff, here with $m=\infty$ in the case $N=1,2$ we find $(\nu_{N=1},\nu_{N=2})=(0.504,0.455)$ and $(\omega_{N=1},\omega_{N=2})=(0.326,0.005)$, these values are very distant from the classical case. 

For $N=2$ the critical exponent $\theta_2$ disappears and "blend" with $\theta_3$ for $m\le11/2$ where they form a couple of complex conjugate pairs. These complex conjugate pairs disappear for $m>11/2$ to form two real different critical exponents. This behavior is obtained both from the shooting and pseudospetral solutions but also independently from the results of the polynomial truncation around the minimum of $u_*$, so we do not believe this is an artifact of the numerical technique. 

The asymptotics of perturbations are of the same form of the perturbations in the C-type cutoff. The explicit expressions are given in Appendix \ref{applinasin}.

%%%%%%%%%%%%%%%%%%%%%%
\section{Conclusions}\label{conc}
%%%%%%%%%%%%%%%%%%%%%%

\noindent In this paper we considered the most general theory coupling the Einstein theory non-minimally with a $O(N)$ set of scalar fields with the aim, on one hand, of testing the Wilsonian proper time functional RG framework and, on the other hand, of confirming the previous results in \cite{Percacci:2015wwa, Labus:2015ska, Borchardt:2015rxa} and enlarge them. To that end we discussed scaling solutions and their critical properties. In $d=4$ the space of fixed-point solutions is composed only by the matter-coupled Reuter-like fixed-point, eq. (\ref{GFP}), and a generalization of this where $f_*$ depends linearly on $x$ (quadratically in the scalar fields), eq. (\ref{scal1}). The gaussian fixed-point exists for both the schemes with $\epsilon=0$ and $\epsilon=1$, whereas its generalization only for $\epsilon=0$. Both fixed-points present a $d$-dependent critical value for the number of fields $N$, where $f_*$ becomes zero and then takes a negative value.

The linearization of the flow equations around the analytic scaling solutions in $d=4$ shows that the gaussian fixed-point gives rise to two sets of spectra for the critical exponents, eq. (\ref{CritExpGFP}), and eigenfunctions (critical directions). One set is the standard result of scalar field theory, the second set is exclusive of the coupling gravity plus matter. This latter set exists only for $\epsilon\ne0$. The well-known standard set of critical exponents does not depend on the number of fields and always gives two relevant and one marginal direction. The new set depends on $N$ and accordingly the number of relevant directions depends on $N$ too. If $N<16$ the number of relevant directions is determined by the closest integer number to $\frac{9}{16-N}+1$. If $16<N<25$ there are no relevant directions. If $N>25$ there is only one relevant direction contained in $j=0$. The total dimension of the UV-critical manifold is given by $2$ plus the number of relevant directions of the new set.

In $d=3$ at finite values of $N$ the space of fixed-point solutions includes all analytic scaling solutions discussed in \ref{secAS} but also a non-trivial scaling solution that is the gravitational-dressed version of the Wilson-Fisher fixed-point. This also yields a non-trivial solution for $f_*(x)$. The properties of these fixed-points depend on $\epsilon$. The solution exists for $N\lesssim9$ and $N\gtrsim8184$ in the C-scheme and for $N\lesssim2$ and $N\gtrsim10000$ in the B-scheme. In the case $\epsilon=0$, the fixed-point potential is not significantly affected by gravity. On the contrary, $\epsilon=1$ has a large effect on the potential and some similarity to the classical Wilson-Fisher solution is lost. The difference reflects also on the solution for $f_*(x)$. With $\epsilon=0$, $f_*(x)$ has a negative first derivative for all values of $x$ so at some point $f_*(x)$ becomes negative. In contrast, with $\epsilon=1$ $f_*(x)$ has a positive first derivative for all values of $x$. For $N=1$ and $N=2$ this is the same behavior observed in \cite{Borchardt:2015rxa} and~\cite{Labus:2015ska}, respectively.

The critical properties of the gravitational-dressed Wilson-Fisher solution for general $N$ depend strongly on $\epsilon$. With $\epsilon=0$ and for all values of $N$ we find only one non-trivial relevant direction whose value is very close to the classical case, while with $\epsilon=1$ we find three non-trivial relevant directions, where the value of the first is very distant from the classical case. For $N=1$ this is the same conclusion obtained in \cite{Borchardt:2015rxa}, although our critical exponents differ a little bit from their results. 

A different situation occurs for the critical exponent $\omega$. Whereas $\nu$ is always a real number, $\omega$ acquires an imaginary part when $\epsilon=0$. We find that this situation is realized for $N=1$ and $N=2$, for $N>2$ the critical exponents become real again. The real part of $\omega$ is very different from the classical value. The same feature was observed in \cite{Percacci:2015wwa} and in particular with $N=1$ we get the same results. With $\epsilon=1$, $\omega$ is real but its value is very distant from the classical case.

The limit $N\to\infty$ is the only case where $\epsilon$ do not play any role in the dynamics. In this limit gravity does not have any influence on the potential $u_*(x)$, which is the same as the flat spacetime case. The fixed-point solutions and the critical exponents can be computed analytically for every value of $d$. We find two lines of fixed-point solutions, which are labeled by two free parameters $c_1$ and $c_2$, eq.(\ref{solanN}). For $d<4$ we find only one admissible non-trivial fixed-point solution, which coincides with the case $(c_1,c_2)=(0,0)$, whose analytical form is given in eq.~\eqref{scalNinf}. These scaling solutions have a gravitational interaction characterized by a simple linear function $f_*(x)$ contrary to the outcome of the analysis for the effective average action. In $d=4$ only the gaussian fixed-point remains. Focusing on the spectrum of the linearized operator around the scaling solution, in contrast to the scaling solutions, the critical exponents do not differentiate between $d=4$ and $d\ne4$, the results are given by eq.~\eqref{critNinf}, which reproduces the well-known results. 

From the physical point of view the outcome of this investigation confirms that in general the presence of matter significantly modifies the structure of the UV-critical manifold, potentially allowing for different continuum limits. Indeed besides the Gaussian fixed point (GFP, \ref{GFP}), two additional fixed points (\ref{gfp1} and \ref{gfp2}) emerge, as already discussed in~\cite{Labus:2015ska} in the EAA (1PI) framework.
Also in $d=3$, the critical properties of the so-called "gravitational" Wilson-Fisher fixed point closely resemble those obtained in the EAA FRG analysis. However, our results extend previous findings by exploring a broader $N$-dependence and regulator dependence (via the parameter $m$). A natural next step would be to incorporate the flow of a field-dependent wavefunction renormalization function, that is to the full second order in the derivative expansion of the scalar sector, which we plan to address in future work.

\newpage
\begin{appendices}

\renewcommand\thesubsection{\arabic{subsection}}
%%%%%%%%%%%%%%%%%%

%%%%%%%%%%%%%%%%%%%%%%%%%%%%%%%%%%%%%%%%%%%%%%%%%%%%%%
\section{Wilsonian flows: coarse graining as a field redefinition}\label{appendix0}

\noindent We collect here a few standard facts~\cite{Wegner:1974sla,Rosten:2010vm} about the UV-regulated Wilsonian renormalization in a continuum formulation. A Wilsonian renormalization-group step consists of two operations: (i) a coarse-graining (or blocking) transformation and (ii) a rescaling. In this appendix we focus on the first (and less trivial) step, which can be implemented in two ways:  on one hand by a direct traditional Kadanoff's blocking procedure; one the other, as suggested by Wegner, by some suitable effective transformation of the field variable which ensures the invariance of the partition function in terms of a transformed action. Let us give a short description of both cases.

\subsection*{Blocking transformation and Wilsonian action}
In a path-integral description, coarse graining is controlled by a UV scale $\Lambda$. A coarse-grained field $\phi$ at scale $\Lambda$ may be defined in terms of a bare field $\phi_0$ (at a microscopic scale $\Lambda_0$) through a blocking map
\begin{equation}
\phi(x)=b^\Lambda[\phi_0](x)\,,
\end{equation}
which in general can be non-linear and mildly non-local, as expected for a reasonable coarse-graining procedure. The associated Wilsonian action $S_\Lambda[\phi]$ is defined by
\begin{equation}
\label{WFdefac}
e^{-S_\Lambda[\phi]}=\int [\mathrm{d}\phi_0]\,
\delta\!\left(\phi-b^\Lambda[\phi_0]\right)\,e^{-S_{\Lambda_0}[\phi_0]}\,.
\end{equation}
By construction, the partition function
\begin{equation}
Z=\int [\mathrm{d}\phi]\,e^{-S_\Lambda[\phi]}
\end{equation}
is invariant under changes of $\Lambda$, i.e. $\Lambda\partial_\Lambda Z=0$.

Different blocking maps $b^\Lambda$ may lead to the same coarse-grained physics, reflecting the fact that coarse graining is, in practice, not invertible and may admit multiple equivalent mathematical realizations. Nevertheless in the continuum limit, if one employs suitable coarse-graining schemes, one can construct RG flow equations that can be solved to give an evolution towards the IR and also the UV (inverted) direction.

\subsection*{General form of Wilsonian flow and definition of \texorpdfstring{$\psi^\Lambda$}{N}}
Differentiating the definition of $S_\Lambda$ w.r.t
%with respect to 
$\Lambda$ and using standard %functional 
manipulations, one finds
\begin{equation}
\label{general_b_flow}
-\Lambda \partial_\Lambda e^{-S_\Lambda[\phi]}
=\int \mathrm{d}x \,\frac{\delta}{\delta \phi(x)}
\left[
\int [\mathrm{d}\phi_0]\,
\delta\!\left(\phi-b^\Lambda[\phi_0]\right)\,
\Lambda\partial_\Lambda b^\Lambda[\phi_0](x)\,
e^{-S_{\Lambda_0}[\phi_0]}
\right]\,.
\end{equation}
This motivates the (implicit) definition of an infinite-dimensional vector of functionals $\psi_x^\Lambda[\phi]$ by
\begin{equation}
\label{general_psi_flow}
-\Lambda \partial_\Lambda e^{-S_\Lambda[\phi]}
\equiv \int \mathrm{d}x\,\frac{\delta}{\delta\phi(x)}
\left(\psi_x^\Lambda[\phi]\,e^{-S_\Lambda[\phi]}\right)\,.
\end{equation}
Knowledge of $\psi_x^\Lambda[\phi]$ is sufficient to specify a Wilsonian RG flow~\footnote{We remind that an infinite family of $\psi_x^\Lambda[\phi]$ exists which leads to the same Wilsonian RG flow equation~\cite{Bonanno:2019ukb}.}.
Expanding \eqref{general_psi_flow} yields the standard functional form
\begin{equation}
\label{wilsonianflowpsi}
\Lambda\partial_\Lambda S_\Lambda[\phi]
=\int \mathrm{d}x\left[
\frac{\delta\psi_x^\Lambda[\phi]}{\delta\phi(x)}
-\psi_x^\Lambda[\phi]\frac{\delta S_\Lambda[\phi]}{\delta\phi(x)}
\right]\,.
\end{equation}

\subsection*{Interpretation as an infinitesimal change of variables}
Equation~\eqref{wilsonianflowpsi} can also alternatively be understood as originating from an infinitesimal change of variables in the functional integral~\cite{Wegner:1974sla}. Consider
\begin{equation}
\phi(x)\to \phi'(x)=\phi(x)+\delta t\,\psi_x^\Lambda[\phi]\,,\qquad
\delta t=\frac{\delta\Lambda}{\Lambda}\,.
\end{equation}
Rewriting the partition function in terms of $\phi$ and exponentiating the Jacobian, %contribution
one finds
\begin{equation}
\label{changevar}
\int [\mathrm{d}\phi']\, e^{-S_\Lambda[\phi']}
=\int [\mathrm{d}\phi]\,
\exp\!\left\{-S_\Lambda[\phi]
+\delta t\int \mathrm{d}x\left[
\frac{\delta\psi_x^\Lambda[\phi]}{\delta\phi(x)}
-\psi_x^\Lambda[\phi]\frac{\delta S_\Lambda[\phi]}{\delta\phi(x)}
\right]\right\}
+\mathcal{O}((\delta t)^2)\,,
\end{equation}
which reproduces \eqref{wilsonianflowpsi} and corresponds to the shift $\Lambda\to\Lambda-\delta\Lambda$ in the coarse-graining picture.

As a last remark we observe that the coarse-graining scheme directly defined through $b^\Lambda$ is in general linked in a highly non trivial way, non locally to some degree and certaily non linearly, to the vector of functionals given by $\psi_x^\Lambda$. Therefore, as anticipated, there are two ways to define a UV-regulated Wilsonian RG flow, either via an explicit form of the suitable coarse-graining ($b^\Lambda$) or directly by a suitable field redefinition ($\psi_x^\Lambda$), as proposed by Wegner.

\subsection*{Possible interpretation of a proper-time flow as an exact UV-regulated Wilsonian flow.}
In order to interpret the PT flow equation~\eqref{PTFE} as the flow of a Wilsonian action then one has to show that there exist a coarse-graining scheme (or better possibly a family of them) that leads to the PT flow equation. This can be accomplished by doing reverse engineering at two different levels. The functional equation can first be solved in terms of $\psi_x^\Lambda[\phi]$ comparing Eqs.~\eqref{general_psi_flow} with~\eqref{PTFE},
\begin{equation} 
%\label{eq:generalCG}
 \Lambda \partial_\Lambda e^{-S_\Lambda[\phi]}
 = \int \! d x \, \frac{\delta}{\delta  \phi(x)} \left( \psi^{\Lambda}_{x}[\phi] e^{- S_{\Lambda}[\phi]} \right)\
 = - e^{- S_{\Lambda}[\phi]} \frac{1}{2} \int_{0}^{\infty} \frac{d s}{s}
 {\rm tr}\left[ r_{\Lambda}(s) e^{-s S^{(2)}_{\Lambda}[\phi]} \right] \,,
 \label{defpsi1}
\end{equation}
which has an infinite family of solutions, depending non trivially on the Wilsonian action $S_\Lambda[\phi]$ and all corresponding to the same Wilsonian RG flow. In~\cite{Bonanno:2019ukb} some of us have computed it for a free theory. The result was non trivial and there has not been yet any attempt to analyze it for an interacting theory; certainly one expect it to be rather complicated.~\footnote{ Let us note that for a Wilsonian Polchinski flow, associated to standard two-point function regulators, the form of $\psi_x^\Lambda[\phi]$ is well known (e.g. see~\cite{Rosten:2010vm}) with an affine dependence on $S_\Lambda$. } Once, possibly, the family of $\{\psi_x^{\Lambda}[\phi]\}$ necessary for a PT flow is found, the problem of defining the UV-regulated Wilsonian RG flow in terms of a field change of variable, in the sense of Wegner, would be solved. 

Then, if one wants to find the precise traditional blocking/coarse-graining procedure to be used in the path integral a second reverse step would consist in deriving the precise form of the associated map $b^\Lambda$. This requires to solve a non trivial implicit differential functional equation
\begin{equation}\label{defb}
\psi_x^\Lambda[\phi]\,e^{-S_\Lambda[\phi]}=
\int [\mathrm{d}\phi_0]\,
\delta\!\left(\phi-b^\Lambda[\phi_0]\right)\,
\Lambda\partial_\Lambda b^\Lambda[\phi_0](x)\,
e^{-S_{\Lambda_0}[\phi_0]}
\,,
\end{equation}
as one may see by comparing Eqs.~\eqref{general_b_flow} and~\eqref{general_psi_flow}.

Summarizing, the existence and the properties of a coarse-graining procedure underlying a RG PT flow along the line illustrated above are not yet known and clarifying them,  certainly deserves a separate careful non trivial study, since it would make such a flow exact in the UV-regulated Wilsonian sense. Given that $\psi_x^\Lambda$ or $b^\Lambda$ are expected to depend on the Wilsonian action itsself, it would be probably interesting to devise well defined expansions, which may provide also approximation schemes, that could guide in the construction of the associated Wilsonian field redefinition/coarse-graining procedure.

%%%%%%%%%%%%%%%%%%%%%%%%%%%%%%%%%%%%%%%%%%%%%%%%%%%%
\section{Derivation of the Hessian}\label{appendix1}
\noindent We shall closely follow here what was done in~\cite{Labus:2015ska}. In order to get the hessian of eq.~(\ref{theorysem}) we need to expand the action around a general background to second order in the fluctuation. We use the exponential splitting for the gravitational field, whereas the linear parametrization for the scalar field:
\begin{equation}\begin{split}
&g_{\mu\nu}={\bar{g}}_{\mu\rho}\left(e^h\right)_{\ \nu}^\rho={\bar{g}}_{\mu\rho}\left(\delta_{\ \nu}^\rho+h_{\ \nu}^\rho+\frac{1}{2}h_{\  \sigma}^\rho h_{\ \nu}^\sigma+\ldots\right)={\bar{g}}_{\mu\nu}+h_{\mu\nu}+\frac{1}{2}h_{\mu\lambda}h_{\ \nu}^\lambda+\ldots\\
&\phi_i={\bar{\phi}}_i+\Phi_i
\end{split}\end{equation}
the barred quantities are the background fields, which are coordinate independent. At the end we have to compute 
\begin{equation}
\delta^2\mathcal{L}=\delta^2\left[\sqrt g\left(-F\left(\rho\right)R+\frac{1}{2}\sum_{i=1}^{N}{\phi_i(-\Box)\phi^i}+U\left(\rho\right)\right)\right]
\end{equation}
and the result gives
\begin{equation}
\label{Hess}
S^{\left(2\right)}=S_{hh}^{\left(2\right)}+S_{\Phi\Phi}^{\left(2\right)}+S_{h\Phi}^{\left(2\right)}
\end{equation}
where
\begin{equation}\begin{split}
\label{hesssplit}
&S_{hh}^{\left(2\right)}=\frac{1}{2}\int{d^dx\sqrt{\bar{g}}h_{\mu\nu}\left\{-F\left(\bar{\rho}\right)\Lambda_{\rho\sigma}^{\mu\nu}\Box+{\bar{A}}_{\rho\sigma}^{\mu\nu}+F\left(\bar{\rho}\right)\left[\delta_\rho^\nu{\bar\nabla}^\mu{\bar\nabla}_\sigma-g^{\mu\nu}{\bar\nabla}_\rho{\bar\nabla}_\sigma\right]\right\}h^{\rho\sigma}}\\
&\resizebox{1.1\hsize}{!}{$S_{\Phi\Phi}^{\left(2\right)}=\frac{1}{2}\sum_{a,b=1}^{N}\int{d^dx\sqrt{\bar{g}}\Phi^a\left\{\left[-\Box+2\bar{\rho}U^{\prime\prime}\left(\bar{\rho}\right)+U^\prime\left(\bar{\rho}\right)-\left[2\bar{\rho}F^{\prime\prime}\left(\bar{\rho}\right)+F^\prime\left(\bar{\rho}\right)\right]\bar{R}\right]P_{ab}^R+\left[-\Box+U^\prime\left(\bar{\rho}\right)-F^\prime\left(\bar{\rho}\right)\bar{R}\right]P_{ab}^T\right\}\Phi^b}$}\\
&S_{h\Phi}^{\left(2\right)}=\sum_{a,b=1}^{N}\int{d^dx\sqrt{\bar{g}}\Phi^aP_{ab}^R{\bar{\phi}}_b\sqrt{2\rho}\left\{\frac{1}{2}{\bar{g}}^{\mu\nu}\left[U^\prime\left(\bar{\rho}\right)-F^\prime\left(\bar{\rho}\right)\bar{R}\right]-F^\prime\left(\bar{\rho}\right)\left(-{\bar{R}}^{\mu\nu}+{\bar\nabla}^\mu{\bar\nabla}^\nu-{\bar{g}}^{\mu\nu}\bar\Box\right)\right\}h_{\mu\nu}}
\end{split}\end{equation}
and
\begin{equation}\begin{split}
\label{valgrav}
&K_{\rho\sigma}^{\mu\nu}=\frac{1}{4}\left(\delta_\rho^\mu\delta_\sigma^\nu+\delta_\sigma^\mu\delta_\rho^\nu\right)-\frac{1}{2}{\bar{g}}^{\mu\nu}{\bar{g}}_{\rho\sigma}\\
&A_{\rho\sigma}^{\mu\nu}=\frac{1}{4}\left(\delta_\sigma^\mu\delta_\rho^\nu-{\bar{g}}^{\mu\nu}{\bar{g}}_{\rho\sigma}\right)\left(F\left(\bar{\rho}\right)\bar{R}-U\left(\bar{\rho}\right)\right)+\\
&+F\left(\bar{\rho}\right)\left[-\frac{1}{4}\left({\bar{R}}_\sigma^\nu\delta_\rho^\mu+{\bar{R}}_\sigma^\mu\delta_\rho^\nu+{\bar{R}}_\rho^\nu\delta_\sigma^\mu\right)+{\bar{g}}^{\mu\nu}{\bar{R}}_{\rho\sigma}-\frac{1}{2}\left({\bar{R}}_{\rho\sigma}^{\mu\nu}+{\bar{R}}_{\sigma\rho}^{\mu\nu}\right)\right]
\end{split}\end{equation}
and
\begin{equation}
P_{ab}^R=\frac{{\bar{\phi}}_a{\bar{\phi}}_b}{{\bar{\phi}}^2},\quad \quad P_{ab}^T=\delta^{ab}-P_{ab}^R=\delta^{ab}-\frac{{\bar{\phi}}_a{\bar{\phi}}_b}{{\bar{\phi}}^2}
\end{equation}
are respectively the radial and longitudinal projection tensors for the multiplet of $O(N)$ in the fundamental vector representation. In eq.~(\ref{hesssplit}) the quantities in curly brackets are the components of Hessian of eq.~(\ref{theorysem}).

It is useful to set a specific background metric to simplify eq.~(\ref{valgrav}). We choose a maximally symmetric background metric
\begin{equation}
\bar{R}_{\mu\nu\lambda\rho}=\frac{\bar{R}}{d(d-1)}(g_{\mu\lambda}\bar{g}_{\nu\rho}-\bar{g}_{\nu\lambda}\bar{g}_{\mu\rho}),\quad \quad \bar{R}_{\mu\nu}=\frac{1}{d}\bar{R}\bar{g}_{\mu\nu}\,.
\end{equation}
Furthermore, it is always possible to choose a basis where
\begin{equation}
P_{ab}^R=\frac{{\bar{\phi}}_a{\bar{\phi}}_b}{{\bar{\phi}}^2}=\left\{\begin{matrix}0&a\neq N,b\neq N\\\frac{{\bar{\phi}}_N^2}{{\bar{\phi}}^2}=1&a=N,b=N\\\end{matrix}\right.,\quad \quad P_{ab}^T=\delta^{ab}-P_{ab}^R=\left\{\begin{matrix}0&a\neq N,b\neq N\\0&a=N,b=N\\1&a=b\neq N\\\end{matrix}\right.
\end{equation}
in this way we get 
\begin{equation}
A_{\rho\sigma}^{\mu\nu}=\frac{1}{4}\left(2\delta_\rho^\mu\delta_\sigma^\nu-{\bar{g}}^{\mu\nu}{\bar{g}}_{\rho\sigma}\right)\left[F\left(\bar\rho\right)R-U\left(\bar\rho\right)\right]+F\left(\bar\rho\right)\frac{d-2}{d{\left(d-1\right)}}\left({\bar{g}}^{\mu\nu}{\bar{g}}_{\rho\sigma}-\delta_\rho^\mu\delta_\sigma^\nu\right)\bar{R}
\end{equation}
and
\begin{equation}\begin{split}
&S_{\Phi\Phi}^{\left(2\right)}+S_{h\Phi}^{\left(2\right)}=\frac{1}{2}\sum_{a=1}^{N-1}\int{d^dx\sqrt{\bar{g}}\Phi^a\left[-\bar \Box+U^\prime\left(\bar{\rho}\right)-F^\prime\left(\bar{\rho}\right)\bar{R}\right]\Phi^a}+\\
&+\frac{1}{2}\int{d^dx\sqrt{\bar{g}}\Phi^N\left[-\Box+2\bar{\rho}U^{\prime\prime}\left(\bar{\rho}\right)+U^\prime\left(\bar{\rho}\right)-\left[2\bar{\rho}F^{\prime\prime}\left(\bar{\rho}\right)+F^\prime\left(\bar{\rho}\right)\right]\bar{R}\right]\Phi^N}+\\
&+\int{d^dx\sqrt{\bar{g}}\Phi^N\sqrt{2\bar{\rho}}\left\{\frac{1}{2}{\bar{g}}^{\mu\nu}\left[U^\prime\left(\bar{\rho}\right)-F^\prime\left(\bar{\rho}\right)\bar{R}\right]-F^\prime\left(\bar{\rho}\right)\left(-{\bar{R}}^{\mu\nu}+{\bar\nabla}^\mu{\bar\nabla}^\nu-{\bar{g}}^{\mu\nu}\bar\Box\right)\right\}h_{\mu\nu}}\,.
\end{split}\end{equation}
This expression shows that the part of the hessian coming from the scalar fields split in a contribution due to the longitudinal fields, or Goldstone bosons, $\Phi^a$ and one due to the transverse field $\Phi^N$. The two contributions do not interact each other and only the transverse field interacts with gravity. 

Eq.~(\ref{Hess}) is too complicate to use for actual computations in the flow equation, to overcome this difficulty we use the York decomposition to partially diagonalize the kinetic operator
\begin{equation}
h_{\mu\nu}=h_{\mu\nu}^T+{\bar\nabla}_\mu\epsilon_\nu+{\bar\nabla}_\nu\epsilon_\mu+{\bar\nabla}_\mu{\bar\nabla}_\nu\sigma-\frac{1}{d}{\bar{g}}_{\mu\nu}\bar\Box\sigma+\frac{1}{d}{\bar{g}}_{\mu\nu}h \,,
\end{equation}
where $h_{\mu\nu}^T$ is the spin $2$ transverse and traceless tensor, $\epsilon_\mu$ is the spin $1$ transverse vector component, $\sigma$ and $h$ are spin $0$ scalars. After long and tedious algebra we get the following result
\begin{equation}\begin{split}
&S^{\left(2\right)}=\frac{1}{2}\int{d^dx\sqrt{\bar{g}}}\bigg\{\frac{1}{2}F\left(\bar{\rho}\right)h_{\mu\nu}^T\left[-{\bar{\Box}}+\frac{2}{d\left(d-1\right)}\bar{R}\right]h^{\mu\nu T}-\frac{\left(d-1\right)\left(d-2\right)}{2d^2}F\left(\bar{\rho}\right)\hat{\sigma}\left(-{\bar{\Box}}\right)\hat{\sigma}-\\
&-\frac{\left(d-1\right)\left(d-2\right)}{2d^2}F\left(\bar{\rho}\right)h\left[-{\bar{\Box}}+\frac{d-2}{2\left(d-1\right)}\bar{R}\right]h+\frac{U\left(\bar{\rho}\right)}{4}h^2-\\
&-\frac{\left(d-1\right)\left(d-2\right)}{d}F\left(\bar{\rho}\right)\ h\sqrt{-{\bar{\Box}}}\sqrt{-\bar\Box-\frac{\bar{R}}{d-1}}\hat{\sigma}\bigg\}+\frac{1}{2}\sum_{a=1}^{N-1}\int{d^dx\sqrt{\bar{g}}\Phi^a\left[-\bar \Box+U^\prime\left(\bar{\rho}\right)-F^\prime\left(\bar{\rho}\right)\bar{R}\right]\Phi^a}+\\
&+\frac{1}{2}\int{d^dx\sqrt{\bar{g}}\Phi^N\left[-\bar \Box+2\bar{\rho}U^{\prime\prime}\left(\bar{\rho}\right)+U^\prime\left(\bar{\rho}\right)-\left[2\bar{\rho}F^{\prime\prime}\left(\bar{\rho}\right)+F^\prime\left(\bar{\rho}\right)\right]\bar{R}\right]\Phi^N}+\\
&+\int{d^dx\sqrt{\bar{g}}\sqrt{2\rho}}\bigg\{-F^\prime\left(\bar{\rho}\right)\left(1-\frac{1}{d}\right)\Phi^N\sqrt{-\bar\Box}\sqrt{-\bar\Box-\frac{1}{d-1}\bar{R}}\hat{\sigma}+\\
&+\Phi^N\left\{-F^\prime\left(\bar{\rho}\right)\frac{d-1}{d}\left[\bar\Box+\frac{d-2}{2\left(d-1\right)}\bar{R}\right]+\frac{U^\prime\left(\bar{\rho}\right)}{2}\right\}h\bigg\}\, ,
\end{split}\end{equation}
where the variables
\begin{equation}
{\hat{\epsilon}}_\nu\mathrm{=}\sqrt{-\bar\Box-\frac{\bar{R}}{d}}\epsilon_\nu, \quad \quad \hat{\sigma}=\sqrt{-\bar\Box}\sqrt{-\bar\Box-\frac{1}{d-1}\bar{R}}\sigma
\end{equation}
are introduced to remove the Jacobian determinants that arise from York decomposition.

In the previous computation we did not add a gauge fixing term for the gravitational part. We need to choose a gauge to have a well-defined theory. To simplify as much as possible the second variation of the action we choose to work with the “physical gauge” where $\epsilon_\mu=0$ and $h=0$ \cite{Percacci:2015wwa}. Since the Jacobian of the transformation is not unity, the price to pay is that we have to add two ghost terms
\begin{equation}
\int{d^dx}\sqrt{\bar{g}}\left[c_\mu\left(-\bar\Box-\frac{\bar{R}}{d}\right)c^\mu+c\left(-\bar\Box\right)c\right]
\end{equation}
so that we get
\begin{equation}\begin{split}
&S^{\left(2\right)}=\int{d^dx\sqrt{\bar{g}}}\bigg(F\left(\bar{\rho}\right)\left\{\frac{1}{4}h_{\mu\nu}^T\left[-\bar\Box+\frac{2}{d\left(d-1\right)}\bar{R}\right]h^{\mu\nu T}-\frac{\left(d-1\right)\left(d-2\right)}{4d^2}\hat{\sigma}\left(-\bar\Box\right)\hat{\sigma}\right\}+\\
&+c_\mu\left(-\bar\Box-\frac{\bar{R}}{d}\right)c^\mu+c\left(-\bar\Box\right)c\bigg)+\frac{1}{2}\sum_{a=1}^{N-1}\int{d^dx\sqrt{\bar{g}}\mathrm{\Phi}^a\left[-\Box+U^\prime\left(\bar{\rho}\right)-F^\prime\left(\bar{\rho}\right)\bar{R}\right]\mathrm{\Phi}^a}+\\
&\frac{1}{2}\int{d^dx\sqrt{\bar{g}}\mathrm{\Phi}^N\left[-\Box+2\bar{\rho}U^{\prime\prime}\left(\bar{\rho}\right)+U^\prime\left(\bar{\rho}\right)-\left[2\bar{\rho}F^{\prime\prime}\left(\bar{\rho}\right)+F^\prime\left(\bar{\rho}\right)\right]\bar{R}\right]\mathrm{\Phi}^N}+\\
&+\int{d^dx\sqrt{\bar{g}}\left\{-\sqrt{2\rho}F^\prime\left(\bar{\rho}\right)\left(1-\frac{1}{d}\right)\Phi^N\sqrt{-\bar\Box}\sqrt{-\bar\Box-\frac{1}{d-1}\bar{R}}\hat{\sigma}\right\}} \,.
\end{split}\end{equation}
We can diagonalize the scalar sector redefining by a further shift the scalar degree of freedom from the metric $\hat{\sigma}$ as follows
\begin{equation}
\sigma^\prime=\hat{\sigma}+\frac{2d}{d-2}\ \frac{\sqrt{2\rho}F^\prime\left(\bar{\rho}\right)}{F\left(\bar{\rho}\right)}\sqrt{\frac{-\bar\Box-\frac{1}{d-1}\bar{R}}{-\bar\Box}}\Phi^N \,.
\end{equation}
The final result for the second variation of the action is
\begin{equation}\begin{split}
\label{Hessdiag}
&S^{\left(2\right)}=\int{d^dx\sqrt{\bar{g}}}\bigg(F\left({\bar{\rho}}\right)\left\{\frac{1}{4}h_{\mu\nu}^T\left[-\bar\Box+\frac{2}{d\left(d-1\right)}\bar{R}\right]h^{\mu\nu T}-\frac{\left(d-1\right)\left(d-2\right)}{4d^2}\sigma^\prime\left(-\bar\Box\right)\sigma^\prime\right\}+\\
&+c_\mu\left(-\bar\Box-\frac{\bar{R}}{d}\right)c^\mu+c\left(-\bar\Box\right)c\bigg)+\frac{1}{2}\sum_{a=1}^{N-1}\int{d^dx\sqrt{\bar{g}}\Phi^a\left[-\Box+\frac{1}{\bar{\rho}}\left[V^\prime\left(\bar{\rho}\right)-F^\prime\left(\bar{\rho}\right)\bar{R}\right]\right]\Phi^a}+\\
&+\int{\resizebox{1.05\hsize}{!}{$d^dx\sqrt{\bar{g}}\Phi^N\frac{1}{2}\left(1+4\bar{\rho}\frac{d-1}{d-2}\frac{\left[F^\prime\left(\bar{\rho}\right)\right]^2}{F\left(\bar{\rho}\right)}\right)\left[-\Box-\frac{2\bar{\rho}F^{\prime\prime}\left(\bar{\rho}\right)+F^\prime\left(\bar{\rho}\right)+\frac{4\bar{\rho}}{d-2}\frac{\left[F^\prime\left(\bar{\rho}\right)\right]^2}{F\left(\bar{\rho}\right)}}{1+4\bar{\rho}\frac{d-1}{d-2}\frac{\left[F^\prime\left(\bar{\rho}\right)\right]^2}{F\left(\bar{\rho}\right)}}\bar{R}+\frac{2\bar{\rho}V^{\prime\prime}\left(\bar{\rho}\right)+V^\prime\left(\bar{\rho}\right)}{1+4\bar{\rho}\frac{d-1}{d-2}\frac{\left[F^\prime\left(\bar{\rho}\right)\right]^2}{F\left(\bar{\rho}\right)}}\right]\Phi^N$}}=\\
&=\int{d^d x \sqrt{\bar{g}}\left(h_{\mu\nu}^T S_{h^Th^T}^{\left(2\right)}h^{\mu\nu T}+\sigma^\prime S_{\sigma\sigma}^{\left(2\right)}\sigma^\prime+c_\mu S_{c^\mu c^\mu}^{\left(2\right)}c^\mu+c S_{cc}^{\left(2\right)}c+\sum_{a=1}^{N-1}\Phi^a S_{\mathrm{\Phi}^T\mathrm{\Phi}^T}^{\left(2\right)}\Phi^a+\Phi^N S_{\mathrm{\Phi}^L\mathrm{\Phi}^L}^{\left(2\right)}\Phi^N\right)} \,.
\end{split}\end{equation}

%%%%%%%%%%%%%%%%%%%%%%%%%%%%%%%%%%%%%%%%%%%%%%%%%%%%%%%%%%%%
\section{Derivation of the flow equations}\label{appendix2}
\noindent In this appendix we derive the flow equations for a generic dimension $d$.

Each term of eq.~(\ref{Hessdiag}) gives a separate contribution to eq.~(\ref{PTFE}) and in particular for this reason we have to pay attention to the wavefunction renormalization of each field. For $h_{\mu\nu}^T$ the function $F_\Lambda/4$ acts as $Z_\Lambda$, whereas for $\sigma$ and $\Phi^N$ the corresponding $Z_\Lambda$ are $C_\sigma=\frac{\left(d-1\right)\left(d-2\right)}{4d^2}$ and $F_\Lambda C_\Phi=F_\Lambda\frac{1}{2}\left(1+4\bar{\rho}\frac{d-1}{d-2}\frac{\left[F_\Lambda^\prime\left(\bar{\rho}\right)\right]^2}{F_\Lambda\left(\bar{\rho}\right)}\right)$. For the ghost fields $Z_\Lambda=1$. The resulting propertime flow is given by

\begin{equation}\begin{split}
\label{PTflow}
&\Lambda\partial_\Lambda S_\Lambda=-\frac{1}{2}\int_{0}^{\infty}{\frac{ds}{s}r\left(s,\Lambda^2\frac{F_\Lambda\left(\bar{\rho}\right)}{4}\right)\mathrm{Tr}\left[e^{-s S_{h^Th^T}^{\left(2\right)}}\right]}-\frac{1}{2}\int_{0}^{\infty}{\frac{ds}{s}r\left(s,\Lambda^2F_\Lambda\left(\bar{\rho}\right)C_\sigma\right)\mathrm{Tr}\left[e^{-s S_{\sigma\sigma}^{\left(2\right)}}\right]}+\\
&+\int_{0}^{\infty}{\frac{ds}{s}r\left(s,\Lambda^2\right)\mathrm{Tr}\left[e^{-s S_{c^\mu c^\mu}^{\left(2\right)}}\right]}+\int_{0}^{\infty}{\frac{ds}{s}r\left(s,\Lambda^2\right)\mathrm{Tr}\left[e^{-s S_{cc}^{\left(2\right)}}\right]}-\\
&-\frac{N-1}{2}\int_{0}^{\infty}{\frac{ds}{s}r\left(s,\Lambda^2\right)\mathrm{Tr}\left[e^{-s S_{\mathrm{\Phi}^T\mathrm{\Phi}^T}^{\left(2\right)}}\right]}-\frac{1}{2}\int_{0}^{\infty}{\frac{ds}{s}r\left(s,\Lambda^2 C_\Phi\right)\mathrm{Tr}\left[e^{-s S_{\mathrm{\Phi}^L\mathrm{\Phi}^L}^{\left(2\right)}}\right]} \,.
\end{split}\end{equation}
Now each piece is of the form
\begin{equation}
\label{PTint}
\int{\frac{ds}{s}r\left(s,Z_\Lambda \Lambda^2\right)\mathrm{Tr} \left[e^{-s\left(Az+B\right)}\right]} \,,
\end{equation}
where $z=-\bar\Box$. In this form, as was done in~\cite{Percacci:2015wwa}, the trace can be evaluated using the heat kernel expansion
\begin{equation}
{\rm Tr}_{(s)}[W(z)]=\frac{1}{{(4\pi)}^\frac{d}{2}}\int{d^dx\sqrt g\sum_{n=0}^{+\infty}{B_{2n}^{\left(s\right)}\left(z\right)Q_{\frac{d}{2}-n}\left(W\right)}} \,.
\end{equation}
the background is chosen to be a maximally symmetric spacetime. The numbers $B_{2n}^{\left(s\right)}\left(z\right)$ are the heat kernel coefficients, which are well-known from the literature results \cite{Vassilevich:2003xt, DeWitt:1964mxt} and depend on the spin $s$ of the specific field. The trace ${\rm Tr}_{(s)}$ is the trace of the space of fields on which $\Box$ acts. The “$Q$ functionals” are given by
\begin{equation}
Q_{\frac{d}{2}-n}=\int_{0}^{\infty}{\frac{1}{\Gamma\left(\frac{d}{2}-n\right)}W\left(z\right)z^{\left(\frac{d}{2}-n\right)-1}dz}\,,
\end{equation}
that is the Mellin transform of $W\left(z\right)$.
In a curved spacetime the heat kernel expansion is an expansion in the curvature invariants $R$, $R_{\mu\nu} R^{\mu\nu}$, $R_{\mu\nu\rho\sigma} R^{\mu\nu\rho\sigma}$ etc \cite{Barvinsky:1990up, Barvinsky1987BeyondTS, Barvinsky:1990uq, Barvinsky:1993en, Avramidi:2000bm}. Up to linear order in $R$ one has
\begin{equation}
\label{TrHeat}
{\rm Tr}_{(s)}\left[W{\left(z\right)}\right]=\frac{1}{\left(4\pi\right)^\frac{d}{2}}\int{d^dx\sqrt g\left[b_0^{\left(s\right)}Q_\frac{d}{2}\left(W\right)+b_2^{\left(s\right)}R \, Q_{\frac{d}{2}-1}\left(W\right)\right]}+O\left(R^2\right) \,.
\end{equation}
For our fields in eq.~(\ref{PTflow}) the heat kernel coefficients are given by \cite{Lauscher:2001ya}
\begin{equation}\begin{split}
&b_0^{\left(0\right)}=1, \quad \quad b_2^{\left(0\right)}=\frac{1}{6}\,,\\
&b_0^{\left(1\right)}=d-1, \quad \quad b_2^{\left(1\right)}=\frac{1}{6}\left(d-1\right)-\frac{1}{d},\\
&b_0^{\left(2\right)}=\frac{(d+1)(d-2)}{2}, \quad \quad b_2^{\left(2\right)}=\frac{1}{6}\frac{(d-5)(d+1)(d+2)}{2(d-1)} .
\end{split}\end{equation}
Inserting eq.~(\ref{TrHeat}) in eq.~(\ref{PTint}) and performing the integral over s using the cutoff in eq.~(\ref{cutoff}) we get
\begin{equation}\begin{split}
&I\left(s,A,B\right)\equiv\int{d^dx\sqrt g}\bigg[-\frac{\Lambda^dm^\frac{d}{2}}{2^{d+1}\pi^\frac{d}{2}\left(\frac{B}{A\Lambda^2m}+1\right)^{m-\frac{d}{2}}}\frac{\Gamma{\left(m-\frac{d}{2}\right)}}{\Gamma\left(m\right)}\left(2A+\Lambda\frac{\partial A}{\partial \Lambda}\right)\\
&\times\left(2b_0^{\left(s\right)}\left(z\right)\left(A\Lambda^2m+B\right)-b_2^{\left(s\right)}\left(z\right)A\left(d-2m\right)R\right)\bigg].
\end{split}\end{equation}
Putting this result in eq.~(\ref{PTflow}) we find
\begin{equation}\begin{split}
&\Lambda\partial_\Lambda S_\Lambda=-\frac{1}{2}\ I\left(2,\frac{F_\Lambda}{4},\frac{RF_\Lambda}{2\left(d-1\right)d}\right)-\frac{1}{2}I\left(0,-\frac{\left(d-2\right)\left(d-1\right)F_\Lambda}{4d^2},0\right)+\frac{1}{2}I\left(1,\frac{1}{2},-\frac{1}{2}\frac{R}{d}\right)+\\
&\frac{1}{2}I\left(0,\frac{1}{2},0\right)-\frac{1}{2}I\left(0,\frac{1}{2}\left(1+\frac{4\left(d-1\right)\rho F_\Lambda^{\prime2}}{\left(d-2\right)F_\Lambda}\right),\frac{1}{2}\left(-\left(2\rho F_\Lambda^{\prime\prime}+F_\Lambda^\prime+\frac{4\rho F_\Lambda^{\prime2}}{\left(d-2\right)F_\Lambda}\right)R+U_\Lambda^\prime+2\rho U_\Lambda^{\prime\prime}\right)\right)-\\
&-\frac{N-1}{2}I\left(0,\frac{1}{2},\frac{1}{2}\left(-RF_\Lambda^\prime+U_\Lambda^\prime\right)\right) .
\end{split}\end{equation}
Performing the algebra, selecting the terms in $R^0$ and $R^1$ and comparing with eq.~(\ref{LHSPT}) one finally gets the flow for $F_\Lambda$ and $V_\Lambda$:
\begin{equation}\begin{split}
\label{FlowU}
&\Lambda\partial_\Lambda U_\Lambda=\frac{\Lambda^dm^\frac{d}{2}\Gamma{\left(m-\frac{d}{2}\right)}}{4\left(4\pi\right)^\frac{d}{2}\Gamma{\left(m\right)}}\Bigg(2\left(d-3\right)d+4\left(N-1\right)\left(1+\frac{U_\Lambda'}{\Lambda^2m}\right)^{\frac{d}{2}-m}+\left(d-1\right)d\epsilon\frac{\Lambda\partial_\Lambda F_\Lambda}{F_\Lambda}+\\
&+4\left(1+\frac{\epsilon\left(2\frac{\Lambda\partial_\Lambda F_\Lambda'}{F_\Lambda'}-\frac{\Lambda\partial_\Lambda F_\Lambda}{F_\Lambda}\right)}{2\left(1+\frac{\left(d-2\right)F_\Lambda}{4\left(d-1\right)\bar{\rho}{F_\Lambda'}^2}\right)}\right)\left(1+\frac{\frac{U_\Lambda'}{\Lambda^2}+\frac{2\bar{\rho}U_\Lambda''}{\Lambda^2}}{m+\frac{4\left(d-1\right)m\bar{\rho}F_\Lambda^{\prime2}}{\left(d-2\right)F_\Lambda}}\right)^{\frac{d}{2}-m}\Bigg)\,,
\end{split}\end{equation}
\begin{equation}\begin{split}
\label{FlowF}
&\Lambda\partial_\Lambda F_\Lambda=\frac{\Lambda^{d-2}m^{\frac{d}{2}-1}\left(\frac{d}{2}-m\right)\Gamma{\left(m-\frac{d}{2}\right)}}{24\left(4\pi\right)^\frac{d}{2}\Gamma{\left(m\right)}}\Bigg(2\left(d-3\right)d+72+4\left(N-1\right)\left(1+6F_\Lambda'\right)\left(1+\frac{U_\Lambda'}{\Lambda^2m}\right)^{\frac{d}{2}-m-1}+\\
&+\left(d^2-d-\frac{24}{d}-24\right)\epsilon\frac{\Lambda\partial_\Lambda F_\Lambda}{F_\Lambda}+\\
&+4\left(1+\frac{\epsilon\left(2\frac{\Lambda\partial_\Lambda F_\Lambda^\prime}{F_\Lambda^\prime}-\frac{\Lambda\partial_\Lambda F_\Lambda}{F_\Lambda}\right)}{2\left(1+\frac{\left(d-2\right)F_\Lambda}{4\left(d-1\right)\bar{\rho}{F_\Lambda'}^2}\right)}\right)\left(1+\frac{6\left(2\bar{\rho}F_\Lambda^{\prime\prime}+\frac{4\bar{\rho}{F_\Lambda'}^2}{\left(d-2\right)F_\Lambda\left(\bar{\rho}\right)}+F_\Lambda'\right)}{1+\frac{4\left(d-1\right)\bar{\rho}{F_\Lambda'}^2}{\left(d-2\right)F_\Lambda}}\right)\left(1+\frac{\frac{U_\Lambda'}{\Lambda^2}+\frac{2\rho U_\Lambda''}{\Lambda^2}}{m+\frac{4\left(d-1\right)m\bar{\rho}{F_\Lambda'}^2}{\left(d-2\right)F_\Lambda}}\right)^{\frac{d}{2}-\left(m+1\right)}\Bigg)
\end{split}\end{equation}
A prime indicates the derivative with respect to $\bar \rho$. The beta functions depend on $\Lambda \partial_\Lambda F_\Lambda$ and $\Lambda \partial_\Lambda F_\Lambda^{\prime}$, this is due to the wavefunction renormalization inside the cutoff eq.~(\ref{cutoff}). The presence of $\Lambda \partial_\Lambda F_\Lambda^{\prime}$ does not allow to solve the system algebraically for $\Lambda \partial_\Lambda F_\Lambda$. Only when $\epsilon=0$ these pieces disappear.

%%%%%%%%%%%%%%%%%%%%%%%%%%%%%%%%%%%%%%%%%%%%%%%%%%%
\section{The numerical techniques}\label{appendix3}
\noindent In this appendix we describe the numerical techniques we used to obtain the numerical solutions of the flow equations.

\subsection{The shooting method}
A scaling solutions of eq.~(\ref{betau}) and eq.~(\ref{betaf}) must be defined for all non-negative real values of $x$, therefore it interpolates smoothly between the origin and infinity. For this reason, we use the "shooting to a fitting point method" \cite{10.5555/1403886}, here an inward integration from infinity and an outward integration from the origin are matched at some fitting point where one requires continuity of the functions and of their derivatives.

In the standard shooting technique, the starting values of the numerical integrations are fixed arbitrarily in a range compatible with the approximations one considers. In our case we improve the method requiring that the starting value of the inward integration is considered as a parameter to be found from the numerical solution of the shooting system. This parameter constrains the numerical system so that at the fitting point we match the solutions up to the third derivative.

In the shooting to a fitting point method the boundary conditions for the Cauchy problem of the numerical integrations are determined by the analytic asymptotic behaviors at $x\to0$ and $x\to\infty$. We illustrate the asymptotic solutions using $d=3$ and $m=d/2+1$ but similar results hold for all values of $m$. Near the origin we find
\begin{equation}\begin{split}
&u_*(x\to0)=u_0-\frac{5 x \left(4 N-3 u_0+6 \epsilon \right)}{12 \epsilon -6 u_0}+O(x^2)\\
&f_*(x\to0)=f_0+\frac{x \left(-60 f_0 N+4 \left(26 N \epsilon +72 N-9 \epsilon ^2\right)+36 u_0 \epsilon -9 u_0^2\right)}{54 \left(u_0-2
   \epsilon \right){}^2}+O(x^2)\,,
\end{split}\end{equation}
where $u_0$, $f_0$ are free parameters. 
%{\color{red} Actually, the solution contains also semi-integer powers: $u_{1/2} x^{1/2}+u_{3/2} x^{3/2}+..$ and $f_{1/2} x^{1/2}+f_{3/2} x^{3/2}+..$, where $u_{1/2}$ and $f_{1/2}$ are other two free parameters. However, $u_{1/2}$ and $f_{1/2}$ have to be set to zero because the derivatives of half-integer powers give a singularity at $x=0$ and the Cauchy problem cannot be defined here. This constraint is nothing but the way the $Z_2$ symmetry eliminates around the origin two of the $4$ parameters. This also yields $u_{3/2}=0$, $f_{3/2}=0$ and so on.} \gpv{Why here we are talking of half integer powers? You want to consider non-power series solutions invoking the use of a Fr\"obenius method with a factor $x^a$ in front, saying that $a=1/2$ is permitted but then saying also that $a=1/2$ is not permitted because of the singular behavior? This does not make much sense! I would simply remove this paragraph!}

The asymptotic behavior at $x\to\infty$ is given by
\begin{equation}\begin{split}
\label{seriesinfasin}
&u_*(x\to\infty)=x^3 u_{\infty }+\frac{36 \epsilon }{5 x f_{\infty }}+\frac{\frac{2 \left(8 f_{\infty }+5 N-4\right)}{15 u_{\infty }}+\frac{96 \epsilon  (13
   \epsilon -36)}{125 f_{\infty }^2}}{x^2}+\frac{32 \left(\frac{\epsilon  (13 \epsilon -72) (13 \epsilon -18)}{f_{\infty
   }^3}-\frac{25 (\epsilon +2)}{u_{\infty }}\right)}{375 x^3}+
   \\
   &\frac{-\frac{625 \left(16 f_{\infty } \left(4 f_{\infty }+1\right)+25 N-24\right)}{u_{\infty }^2}-\frac{28800 (\epsilon +1)
   (13 \epsilon -24)}{f_{\infty } u_{\infty }}+\frac{384 \epsilon  (13 \epsilon -108) (13 \epsilon -36) (13 \epsilon
   -12)}{f_{\infty }^4}}{39375 x^4}+O\left(\frac{1}{x^5}\right)\\
&f_*(x\to\infty)=x f_{\infty }+\frac{24}{5}+\frac{104 \epsilon }{25 x
   f_{\infty }}+\frac{416 (13 \epsilon -36) \epsilon }{1125 x^2 f_{\infty }^2}+\frac{208 (13 \epsilon -72) (13 \epsilon -18)
   \epsilon }{5625 x^3 f_{\infty }^3}+\\
   &+\frac{\frac{1664 \epsilon  (13 \epsilon -108) (13 \epsilon -36) (13 \epsilon -12)}{421875 f_{\infty }^4}-\frac{2 f_{\infty }
   \left(152 f_{\infty }+75 N-52\right)+25 N-24}{675 u_{\infty }^2}}{x^4}+O\left(\frac{1}{x^5}\right) \,,
\end{split}\end{equation}
where $u_\infty$ and $f_\infty$ are two free parameters. The powers $x^3$ for $u_*$ and $x$ for $f_*$ are expected from the classical scaling behavior of the flow equations. %Contrary to the solutions near the origin, here there are only two free parameters, which means that only a subset of fixed point solutions of eq.~(\ref{betau}) and eq.~(\ref{betaf}) admits an asymptotic behavior. \gpv{In principle I think it is possible to introduce in a refined asymptotic analysis with exponential and not power like terms a dependence in the other two cauchy parameters, but if I remember they are totally negligible and it is impossible to deal numerically with them. So I am not sure it is good to mention all this...}

We find that for a better numerical stability the first derivatives of eqs.~(\ref{betau}) and (\ref{betaf}) are more suitable for the numerical computations than the standard flow equations. This leads to the study of a system of equations of second order in $v_*=u_*'$ and third order in $f_*$. 

Using for the numerical inward and outward integrations the previous power series and their derivatives as boundary conditions, evaluated respectively at the starting points $x_{min}$ and $x_{max}$, looking for a scaling solution means to find the simultaneous values of $u_0$, $f_0$, $u_\infty$, $f_\infty$ such that the fixed point solution and its derivatives are continuous and differentiable at the fitting point $x_{fit}$. This then ensures the continuity and the differentiability to all other values of $x$. Calling respectively $v_{IN}$, $f_{IN}$, $v_{OUT}$ and $f_{OUT}$ the numerical inward and outward solutions we require
\begin{equation}\begin{split}
\label{sysnum}
&v_{IN}\left(u_0,f_0\right)=v_{OUT}\left(u_\infty,f_\infty,x_{max}\right), \quad f_{IN}\left(u_0,f_0\right)=f_{OUT}\left(u_\infty,f_\infty,x_{max}\right)\,,\\
&v_{IN}^\prime\left(u_0,f_0\right)=v_{OUT}^\prime\left(u_\infty,f_\infty,x_{max}\right), \quad f_{IN}^\prime\left(u_0,f_0\right)=f_{OUT}^\prime\left(u_\infty,f_\infty,x_{max}\right)\,,\\
&f_{IN}^{\prime\prime}\left(u_0,f_0\right)=f_{OUT}^{\prime\prime}\left(u_\infty,f_\infty,x_{max}\right)\,.
\end{split}\end{equation} 
The solution of this system determines the scaling solutions. The numerical integrations and the solution of this system are obtained by \href{https://reference.wolfram.com/language/ref/NDSolve.html}{NDSolve} and \href{https://reference.wolfram.com/language/ref/FindRoot.html}{FindRoot} algorithms of the Mathematica software. The numerical integrations are performed using $x_{MIN}=10^{-10}$ and $x_{fit}=3$, then we set a tolerance of FindRoot solution of $10^{-8}$. The numerical results do not depend on $x_{fit}$ and the same scaling solutions are obtained with other numerical parameters. 

The starting guesses can be obtained from $u_*(x)=\frac{\lambda_*}{8\pi g_*}+u_*^{WF}\left(x\right)$ and $f_*(x)=\frac{1}{16\pi g_*}-b_*\left(x\right)$ in the limits $x\rightarrow0$ and $x\rightarrow\infty$:
\begin{equation}\begin{split}
\label{guess}
&u_\ast\left(x\rightarrow0\right)=\frac{\lambda_\ast^{PG}}{8\pi g_\ast^{PG}}+u_\ast^{WF}\left(x\rightarrow0\right), \quad \quad u_*\left(x\rightarrow\infty\right)=\frac{\lambda_\ast^{PG}}{8\pi g_\ast^{PG}}+u_\ast^{WF}\left(x\rightarrow+\infty\right)\,,\\
&f_\ast\left(x\rightarrow0\right)=\frac{1}{16\pi g_\ast^{PG}}+b_\ast\left(x\rightarrow0\right), \quad \quad f_\ast\left(x\rightarrow+\infty\right)=\frac{1}{16\pi g_\ast^{PG}}-b_\ast\left(x\rightarrow\infty\right). 
\end{split}\end{equation}
The exact values of $b_\ast\left(x\rightarrow\infty\right)$ and $b_\ast\left(x\rightarrow0\right)$ are not important and they can be put to zero. The reason is that FindRoot finds a numerical solution and then this solution can be used as a new guess.

All numerical solutions are found with $x_{max}\ge30$ and $x_{max}\ge120$, respectively, in the C- and B-type cutoff. In the respective cutoff, with these values of the starting point for the inward integration, the truncation in eq.~(\ref{seriesinfasin}) does not affect the numerical integration. %{\color{red} and the only source of error could come from systematic errors}. \gpv{Is this needed?}  

The analytical technique described to obtain the scaling solutions of $N\to\infty$ limit in eq.~(\ref{scalNinf}) is equivalent to solve the system eq.~(\ref{sysnum}), when the general solution of the fixed-point equations is known analytically. Here $x_{max}$ and $x_{min}$ are infinity and zero, the role of the fitting point $x_{fit}$ is played by the point $x_0$ around which we Taylor expand the general solution. The analytical solutions of the system are given by:
\begin{equation}\begin{split}
&u_0=\frac{5}{6 v_0+15}, \quad \quad u_\infty=\frac{(d-2)^{\frac{d}{d-2}} m^{\frac{d}{d-2}} \left(\frac{\Gamma \left(m-\frac{d}{2}\right)}{\Gamma \left(2-\frac{d}{2}\right)\Gamma (m)}\right)^{\frac{2}{d-2}}}{d}\,,\\
&f_0=-\frac{1}{6},\quad \quad f_\infty=-\frac{1}{6}\,,
\end{split}\end{equation}
where $v_0$ is given by eq.~(\ref{solv0Ninf}).

\subsection{The pseudospectral method}
In the pseudospectral method, which was used in \cite{Borchardt:2015rxa}, a split of the range $[0,+\infty]$ in $[0,+x_0]\cup[x_0,+\infty]$ had to be considered, and a compactification in $[x_0,+\infty]$ was introduced with a parameter $L$. The values of $x_0$ and $L$ have been chosen arbitrarily. In this way, the pseudospectral system matches in $x_0$ only the function and its first derivative. This constraint is not strong enough to obtain a good numerical solution for a higher derivative system because at $x_0$ the second derivatives will have a jump or some other sort of discontinuity. Furthermore, the arbitrarily $x_0$ and $L$ chosen affect the convergence of the solutions. To remove possible discontinuities from the second derivatives, to obtain better convergence and a better numerical solution, we do not fix $x_0$ and $L$ arbitrarily, but we consider them as further parameters of the pseudospectral system, requiring as a further condition the matching of the second derivatives at $x_0$.

In contrast to the shooting method, here we use eq.~(\ref{betaf}) and the first derivative of eq.~(\ref{betau}) as numerical system. The pseudospectral solution of this system is obtained with the collocation method described in \cite{Borchardt:2015rxa}. For the collocation points, we use the Gauss-Lobatto grid \cite{article}.

To apply the collocation method, we decompose $f_*$ and the first derivative $v_*$ of the potential as a sum of two series of Chebyshev polynomials:
\begin{equation}\begin{split}
\label{PSans}
&v_*(x)=v_{B}(x) H (x_0-x)+v_{U}(x) H (x-x_0) \, ,\\
&f_*(x)=f_{B}(x) H (x_0-x)+f_{U}(x) H (x-x_0) \,,
\end{split}\end{equation}
where $H(x)$ is the Heaviside function and
\begin{equation}\begin{split}
&v_{B}(x)=\sum _{i=0}^p c^{(v)}_i T_i\left(\frac{2 x}{x_0}-1\right), \quad \quad v_{U}(x)=x^{\frac{d}{d-2}-1} \sum _{i=0}^p r^{(v)}_i T_i\left(\frac{x-x_0-L}{x-x_0+L}\right)\,,\\
&f_B(x)=\sum _{i=0}^p c^{(f)}_i T_i\left(\frac{2 x}{x_0}-1\right), \quad \quad f_U(x)=x \sum _{i=0}^p r^{(f)}_i T_i\left(\frac{x-x_0-L}{x-x_0+L}\right)\,.
\end{split}\end{equation}
$c^{(v)}_i$, $c^{(f)}_i$, $r^{(v)}_i$ and $r^{(f)}_i$ are the coefficients that the pseudospectral method determines, $p$ is the order of truncation, and $T_i(x)$ is the Chebyshev polynomial of order $i$. 

The numerical system to solve is given by
\begin{equation}\begin{split}
\label{sysPS}
&\beta_v \left(x_i^{B},v_B(x_i^{B}),f_B(x_i^{B})\right)=0, \quad \quad \beta_f \left(x_i^{B},v_B(x_i^{B}),f_B(x_i^{B})\right)=0, \quad \quad i=1,\ldots, p+1\\
&\beta_v \left(x_j^{U},v_U(x_i^{U}),f_U(x_j^{U})\right)=0, \quad \quad \beta_f \left(x_j^{U},v_U(x_j^{U}),f_U(x_j^{U})\right)=0, \quad \quad j=1,\ldots, p-1\\
&v_B(x_0)=v_U(x_0), \quad \quad v_B'(x_0)=v_U'(x_0), \quad \quad v_B''(x_0)= v_U''(x_0)\\
&f_B(x_0)=f_U(x_0), \quad \quad f_B'(x_0)=f_U'(x_0),\quad \quad f_B''(x_0)=f_U''(x_0)\\
\end{split}\end{equation}
where $x_i^{B}$ and $x_i^{U}$ are the collocation points determined by the Gauss-Lobatto grid:
\begin{equation}
-\cos\left(\frac{i\pi}{p}\right)=\frac{2 x_i^{B}}{x_0}-1,\quad \quad -\cos\left(\frac{j\pi}{p}\right)=\frac{x_j^{U}-x_0-L}{x_j^{U}-x_0+L}
\end{equation}
The system eq.(\ref{sysPS}) for the Chebyshev coefficients, $x_0$ and $L$ is made up of algebraic equations. %To remove as many spurious solutions as possible,% 
We set a FindRoot solution accuracy of $10^{-64}$. For true scaling solutions, our improved pseudospectral method determines, respectively, $(L,x_0)=(3,4)$ and $(L,x_0)=(82,165)$ for the C- and B-type cutoff. 

\section{The linearized system}\label{appendix4}
\noindent In this appendix, we give the main equations and describe the numerical techniques for studying the spectrum of the linearized system in $d=3$ around a fixed point.

\subsection{The linearized equations}
The linearized equation for $\delta u$ in $d=3$ is given by
\begin{equation}\begin{split}
\label{eqlinu}
&0=(3-\theta )\delta u+\\
&\resizebox{1.05\hsize}{!}{$+\left(\frac{2 (2 m-3) (N-1)
   \left(\frac{m+u'}{m}\right)^{\frac{3}{2}-m}}{m+u'}-x+Z^{\frac{3}{2}-m} \left(\frac{2 f (2 m-3)}{8 m x \left(f'\right)^2+f \left(m+2 x
   u''+u'\right)}+\frac{8 (3-2 m) x \epsilon  f' \left(2 f x f''+f' \left(f-x f'\right)\right)}{\left(8 x
   \left(f'\right)^2+f\right) \left(8 m x \left(f'\right)^2+f \left(m+2 x u''+u'\right)\right)}\right)\right)\delta u'+$}\\
&+Z^{\frac{5}{2}-m} \left(\frac{4 f m (2 m-3) x \left(8 x \left(f'\right)^2+f\right)}{\left(8 m x
   \left(f'\right)^2+f \left(m+2 x u''+u'\right)\right)^2}+\frac{16 (3-2 m) m x^2 \epsilon  f' \left(2 f x f''+f' \left(f-x
   f'\right)\right)}{\left(8 m x \left(f'\right)^2+f \left(m+2 x u''+u'\right)\right)^2}\right)\delta u''+\\
&\resizebox{1.05\hsize}{!}{$+\Bigg(\frac{\epsilon  \left(6 f \theta -6
   x f'\right)}{f^2}+Z^{\frac{1}{2}-m} \left(\frac{16 (2 m-3) x \left(f'\right)^2 \left(2 x u''+u'\right)}{m \left(8 x
   \left(f'\right)^2+f\right)^2}+\frac{64 (3-2 m) x^2 \epsilon  \left(f'\right)^3 \left(2 f x f''+f' \left(f-x
   f'\right)\right) \left(2 x u''+u'\right)}{f m \left(8 x \left(f'\right)^2+f\right)^3}\right)-$}\\
   &- Z^{\frac{3}{2}-m}\frac{16 x \epsilon  f'
   \left(2 f^2 x f''+f' \left(f^2 (\theta +1)-2 x f' \left(-4 f \theta  f'+4 x
   \left(f'\right)^2+f\right)\right)\right)}{f^2 \left(8 x \left(f'\right)^2+f\right)^2}\Bigg)\delta f+\\
&\resizebox{1.05\hsize}{!}{$+\Bigg(\frac{6 x \epsilon }{f}+Z^{\frac{1}{2}-m} \left(\frac{32 f (3-2 m) x f' \left(2 x u''+u'\right)}{m \left(8 x
   \left(f'\right)^2+f\right)^2}+\frac{128 (2 m-3) x^2 \epsilon  \left(f'\right)^2 \left(2 f x f''+f' \left(f-x
   f'\right)\right) \left(2 x u''+u'\right)}{m \left(8 x \left(f'\right)^2+f\right)^3}\right)+$}\\
   &+\frac{16 x \epsilon 
   Z^{\frac{3}{2}-m} \left(-8 x^2 \left(f'\right)^4+f x \left(f'\right)^2 \left(-16 x f''+16 \theta  f'-3\right)+2 f^2
   \left(x f''+(\theta +1) f'\right)\right)}{f \left(8 x \left(f'\right)^2+f\right)^2}\Bigg)\delta f' \,.
\end{split}\end{equation}
The linearized equation for $\delta f$ is given by
\begin{equation}\begin{split}
\label{eqlinf}
&0=\Bigg(1-\theta+\frac{13 (2 m-3) \epsilon  \left(f \theta -x f'\right)}{6 f^2 m}+Z^{\frac{1}{2}-m} \Bigg(\frac{8 (2 m-3) x \left(f'\right)^2 \left(4 x f''+2 f'-1\right)}{m \left(8 x
   \left(f'\right)^2+f\right)^2}-\\
   &-\frac{4 (3-2 m) x \epsilon  f'}{3 f^2 m \left(8 x \left(f'\right)^2+f\right)^3}\big(-256 x^3 \left(f'\right)^6+32 f x^2 \left(f'\right)^4
   \left(8 \theta  f'-3\right)-\\
   &\resizebox{1.05\hsize}{!}{$-2 f^2 x \left(f'\right)^2 \left(96 x^2 \left(f''\right)^2-44 x f''-24 (\theta -1)
   \left(f'\right)^2-2 f' \left(24 (\theta -2) x f''+10 \theta +11\right)+1\right)+f^3 \left(12 x f''+6 f'+1\right) \left(2 x
   f''+(\theta +1) f'\right)\big)\Bigg)+$}\\
   &+Z^{\frac{3}{2}-m} \Bigg(-\frac{4 (3-2 m)
   (3-2 (m+1)) x \left(f'\right)^2 \left(32 x \left(f'\right)^2+f \left(12 x f''+6 f'+1\right)\right) \left(2 x
   u''+u'\right)}{3 \left(8 x \left(f'\right)^2+f\right) \left(8 m x \left(f'\right)^2+f \left(m+2 x
   u''+u'\right)\right)^2}+\\
   &\resizebox{1.05\hsize}{!}{$+\frac{16 (3-2 m) (3-2 (m+1)) x^2 \epsilon  \left(f'\right)^3 \left(f \left(2 x f''+f'\right)-x
   \left(f'\right)^2\right) \left(32 x \left(f'\right)^2+f \left(12 x f''+6 f'+1\right)\right) \left(2 x u''+u'\right)}{3 f
   \left(8 x \left(f'\right)^2+f\right)^2 \left(8 m x \left(f'\right)^2+f \left(m+2 x u''+u'\right)\right)^2}\Bigg)\Bigg)\delta f+$}\\
&\resizebox{1.05\hsize}{!}{$+\Bigg(\frac{13 (2 m-3) x \epsilon }{6 f m}+\frac{2 (2 m-3) (N-1) \left(\frac{m+u'}{m}\right)^{\frac{1}{2}-m}}{m}-x+Z^{\frac{1}{2}-m} \Bigg(-\frac{2 f (3-2 m) Z^{\frac{1}{2}-m} \left(f-8 x f' \left(4 x f''+f'-1\right)\right)}{m \left(8 x\left(f'\right)^2+f\right)^2}+$}\\
&+\frac{4 (3-2 m) x \epsilon }{3 f m \left(8 x \left(f'\right)^2+f\right)^3}\big(-256 x^3 \left(f'\right)^6-8 f x^2 \left(f'\right)^4 \left(52 x f''-64 \theta  f'+19\right)-\\
&\resizebox{1.05\hsize}{!}{$-f^2 x \left(f'\right)^2 \left(36
   x f'' \left(16 x f''-3\right)-48 (2 \theta -1) \left(f'\right)^2+8 f' \left(-24 (\theta -2) x f''-10 \theta
   -11\right)+3\right)+2 f^3 \left(x f'' \left(12 x f''+1\right)+(6 \theta +9) \left(f'\right)^2+f' \left(12 (\theta +2) x
   f''+\theta +1\right)\right)\Big)\Bigg)+$}\\
   &+Z^{\frac{3}{2}-m} \Bigg(\frac{8 f (2 m-3) (2 m-1)
   x f' \left(12 f x f''+32 x \left(f'\right)^2+6 f f'+f\right) \left(2 x u''+u'\right)}{3 \left(8 x
   \left(f'\right)^2+f\right) \left(8 m x \left(f'\right)^2+f \left(m+2 x u''+u'\right)\right)^2}+\\
   &\resizebox{1.05\hsize}{!}{$+\frac{32 (2 m-3) (2 m-1) x^2 \epsilon  \left(f'\right)^2 \left(-2 f x f''+x \left(f'\right)^2-f
   f'\right) \left(12 f x f''+32 x \left(f'\right)^2+6 f f'+f\right) \left(2 x u''+u'\right)}{3 \left(8 x
   \left(f'\right)^2+f\right)^2 \left(8 m x \left(f'\right)^2+f \left(m+2 x u''+u'\right)\right)^2}\Bigg)\Bigg)\delta f'+$}\\
&+Z^{\frac{1}{2}-m} \left(\frac{4 f (2 m-3) x}{m \left(8 x \left(f'\right)^2+f\right)}-\frac{8 (2 m-3) x^2
   \epsilon  f' \left(24 f x f''+26 x \left(f'\right)^2+12 f f'+f\right)}{3 m \left(8 x \left(f'\right)^2+f\right)^2}\right)\delta f''+\\
&\resizebox{1.05\hsize}{!}{$\Bigg(-\frac{(2 m-3) (2 m-1) (N-1) \left(6 f'+1\right)
   \left(\frac{m+u'}{m}\right)^{\frac{3}{2}-m}}{6 \left(m+u'\right)^2}+Z^{\frac{3}{2}-m} \Bigg(-\frac{f (2 m-3) (2 m-1)
   \left(12 f x f''+32 x \left(f'\right)^2+6 f f'+f\right)}{6 \left(8 m x \left(f'\right)^2+f \left(m+2 x
   u''+u'\right)\right)^2}+$}\\
   &-\frac{2 (2 m-3) (2 m-1) x \epsilon  f' \left(-2 f x f''+x \left(f'\right)^2-f f'\right) \left(12 f x f''+32 x
   \left(f'\right)^2+6 f f'+f\right)}{3 \left(8 x \left(f'\right)^2+f\right) \left(8 m x \left(f'\right)^2+f \left(m+2 x
   u''+u'\right)\right)^2}\Bigg)\Bigg)\delta u'+\\
&+ Z^{\frac{5}{2}-m} \Bigg(-\frac{f m (2 m-3) (2 m-1) x \left(8 x \left(f'\right)^2+f\right) \left(12 f x f''+32 x
   \left(f'\right)^2+6 f f'+f\right)}{3 \left(8 m x \left(f'\right)^2+f \left(m+2 x u''+u'\right)\right)^3}-\\
   &-\frac{4 m (2 m-3) (2 m-1) x^2 \epsilon  f' \left(-2 f x f''+x
   \left(f'\right)^2-f f'\right) \left(12 f x f''+32 x \left(f'\right)^2+6 f f'+f\right)}{3 \left(8 m x \left(f'\right)^2+f
   \left(m+2 x u''+u'\right)\right)^3}\Bigg)\delta u'' \,,
\end{split}\end{equation}
where $Z=\frac{8 m x f'(x)^2+f(x) \left(m+2 x u''(x)+u'(x)\right)}{m \left(8 x f'(x)^2+f(x)\right)}$.

\subsection{Shooting and pseudospectral method for the linearized system}\label{applinasin}
To solve eqs.~(\ref{eqlinu}) and (\ref{eqlinf}) we use again the shooting and the pseudospectral method as a check. 

In the shooting, as numerical system, we use eqs.~(\ref{eqlinu}) and (\ref{eqlinf}). The boundary conditions are given by the power series around $x\to0$ and $x\to\infty$ along with their first derivatives evaluated at $x_{min}$ and $x_{max}$. For illustration we use $m=d/2+1$ but similar results hold for all values of $m$. Near the origin we find 
\begin{equation}\begin{split}
&\delta u(x)=\delta u_0+N x \left(\frac{20 \delta u_0 (\theta -3)}{9 (u_0-2 \epsilon )^2}-\frac{40 \delta f_0 \theta  \epsilon }{3 f_0 (u_0-2 \epsilon )^2}\right)+O(x^2)\,,\\
&\resizebox{1.05\hsize}{!}{$\delta f=\delta f_0+N x \left(\frac{\delta f_0 \left(4 \theta  \epsilon  \left(13 u_0+26 \epsilon +144\right)-30 f_0
   \left((\theta -1) u_0+2 (\theta +1) \epsilon \right)\right)}{27 f_0 \left(2 \epsilon -u_0\right){}^3}-\frac{4 \delta u_0 (\theta -3) \left(15 f_0-26 \epsilon -72\right)}{81 \left(u_0-2 \epsilon \right){}^3}\right)+O(x^2)$}\,,
\end{split}\end{equation}
where $\delta u_0$ and $\delta f_0$ are two free parameters. By linearity we set $\delta u_0=1$ so $\delta u(x=0)=1$. %As for the scaling solutions, smoothness at $x=0$ requires that the coefficients of $x^{n/2+1}$ to zero.\gpv{Same comment as before in pag. 32!}  

In the asymptotic regimes $x\to\infty$ the solution is a superposition of a power series and an exponential. Fixing the coefficients of the exponential term to zero restricts $\theta$ to a discrete set of complex numbers. The power series are given by
\begin{equation}\begin{split}
&\delta u(x\to\infty)=x^{3-\theta}\left(\delta u_{\infty }-\frac{2 \epsilon  \delta f_{\infty}}{x^3 f_{\infty }}+\frac{36 \epsilon  \delta f_{\infty}}{5 x^4 f_{\infty }^2}+\frac{\frac{96 \epsilon  (13 \epsilon -36) \delta f_{\infty}}{125 f_{\infty }^3}-\frac{2
   (\theta -3) \delta u_{\infty} (2 \theta -25 N+20)}{225 u_{\infty }^2}}{x^5}+O\left(\frac{1}{x^6}\right)\right)\,,\\
&\delta f(x\to\infty)=x^{1-\theta}\Bigg(\delta f_{\infty}-\frac{26 \epsilon 
   \delta f_{\infty}}{15 x f_{\infty }}+\frac{104 \epsilon  \delta f_{\infty}}{25 x^2 f_{\infty }^2}+\frac{416 \epsilon  (13 \epsilon -36) \delta f_{\infty}}{1125 x^3 f_{\infty }^3}+\frac{208 \epsilon  \left(169 \epsilon ^2-1170 \epsilon +1296\right)
   \delta f_{\infty}}{5625 x^4 f_{\infty }^4}+\\
   &+\frac{\frac{1664 \epsilon  \left(2197 \epsilon ^3-26364
   \epsilon ^2+73008 \epsilon -46656\right) \delta f_{\infty}}{421875 f_{\infty }^5}-\frac{2 (\theta -1)
   \delta f_{\infty} (2 \theta -25 n+24)}{225 u_{\infty }^2}+\frac{2 (\theta -3) \delta u_{\infty} (2
   (\theta +60)-125 n)}{10125 u_{\infty }^3}}{x^5}+O\left(\frac{1}{x^6}\right)\Bigg) \,,
\end{split}\end{equation}
\begin{comment}
   \frac{32 \epsilon  \left(169 \epsilon
   ^2-1170 \epsilon +1296\right) \delta f_{\infty}}{375 x^6 f_{\infty }^4}
   
   \frac{\frac{1664 \epsilon  \left(28561 \epsilon ^4-507507 \epsilon ^3+2454894 \epsilon ^2-3891888
   \epsilon +1679616\right) \text{$\delta $f}_{\infty }}{3796875 f_{\infty }^6}-\frac{13 \epsilon  \text{$\delta $f}_{\infty }
   \left(4 \left(-\left(3 \theta ^2+63 \theta +170\right) f_{\infty }+76 f_{\infty }^2-6\right)+25 n \left(6 (\theta +5)
   f_{\infty }+1\right)\right)}{30375 f_{\infty }^2 u_{\infty }^2}+\frac{104 (\theta -3) \epsilon  \text{$\delta $u}_{\infty }
   (2 (\theta +60)-125 n)}{455625 f_{\infty } u_{\infty }^3}}{x^6}
\end{comment}
where also $\delta u_\infty$ and $\delta f_\infty$ are two free parameters. Conversely to the solutions near the origin, here the exponents depend on $3-\theta$ and $1-\theta$, this is expected from the scaling argument.

The shooting to a fitting point requires the matching:
\begin{equation}\begin{split}
\label{sysnumlin}
&\delta u_{IN}\left(\delta f_0,\theta\right)=\delta u_{OUT}\left(\delta u_\infty,\delta f_\infty,\theta\right),\ \ \delta f_{IN}\left(\delta f_0,\theta\right)=\delta f_{OUT}\left(\delta u_\infty,\delta f_\infty,\theta\right)\\
&\delta u_{IN}^\prime\left(\delta f_0,\theta\right)=\delta u_{OUT}^\prime\left(\delta u_\infty,\delta f_\infty,\theta\right),\ \ \delta f_{IN}^\prime\left(\delta f_0,\theta\right)=\delta f_{OUT}^\prime\left(\delta u_\infty,\delta f_\infty,\theta\right)
\end{split}\end{equation} 
the solutions fix $\delta f_0$, $\delta u_\infty$, $\delta f_\infty$ and $\theta$ uniquely. For the computation we set an accuracy of the findroot solution of $10^{-8}$.

In the pseudospectral method, as numerical system, we use eq.~(\ref{eqlinf}) and the first derivative of eq.~(\ref{eqlinu}). This is a second
order system with respect to $\delta v =\delta u'$ and $\delta f$. The Chebyshev expansions are given by
\begin{equation}\begin{split}
\label{PSanslin}
&\delta v(x)=\delta v_{B}(x) H (x_0-x)+\delta v_{U}(x) H (x-x_0)\\
&\delta f(x)=\delta f_{B}(x) H (x_0-x)+\delta f_{U}(x) H (x-x_0)
\end{split}\end{equation}
where
\begin{equation}\begin{split}
&\delta v_{B}(x)=\sum _{i=0}^p \delta c^{(v)}_i T_i\left(\frac{2 x}{x_0}-1\right), \quad \quad \delta v_{U}(x)=x^{3-\theta} \sum _{i=0}^p \delta r^{(v)}_i T_i\left(\frac{x-x_0-L}{x-x_0+L}\right)\\
&\delta f_B(x)=\sum _{i=0}^p \delta c^{(f)}_i T_i\left(\frac{2 x}{x_0}-1\right), \quad \quad \delta f_U(x)=x^{1-\theta} \sum _{i=0}^p \delta r^{(f)}_i T_i\left(\frac{x-x_0-L}{x-x_0+L}\right)
\end{split}\end{equation}
$\delta c^{(v)}_i$, $\delta c^{(f)}_i$, $\delta r^{(v)}_i$ and $\delta r^{(f)}_i$ are coefficients determined by
\begin{equation}\begin{split}
&\delta \beta_v \left(x_i^{B},\delta v_B(x_i^{B}),\delta f_B(x_i^{B})\right)=0, \quad \quad \delta\beta_f \left(x_i^{B},\delta v_B(x_i^{B}),\delta f_Bx_i^{B})\right)=0, \quad \quad i=1,\ldots, p+1\\
&\delta \beta_v \left(x_i^{U},\delta v_U(x_i^{U}),f_U(x_i^{U})\right)=0, \quad \quad \delta\beta_f \left(x_i^{U},\delta v_U(x_i^{U}),\delta f_U(x_i^{U})\right)=0, \quad \quad i=1,\ldots, p-1\\
&\delta v_B(0)=1,\quad \quad \delta v_B(x_0)=\delta v_U(x_0), \quad \quad \delta v_B'(x_0)=\delta v_U'(x_0),\quad \quad \delta v_B'(x_0)=\delta v_U'(x_0)\\
&\delta f_B(x_0)=\delta f_U(x_0), \quad \quad \delta f_B'(x_0)=\delta f_U'(x_0),\quad \quad \delta f_B''(x_0)=\delta f_U''(x_0)\\
\end{split}\end{equation}
the condition $\delta v_B(0)=1$ sets the normalization for $\delta v$. As for the scaling solutions, we set an accuracy of findroot solution of $10^{-64}$. %\gpv{when using the FindRoot search within Mathematica} {\color{red} to remove as many spurious solutions as possible} \gpv{In genere aumentare la precisione non garantisce di diminuire la presenza di soluzioni spurie!!!}.

\subsection{The polynomial truncation around the minimum of \texorpdfstring{$u_*$}{N}}\label{appendixPolTr}
Solutions of the flow equations by polynomial truncations are useful to obtain an approximate stability matrix from which the critical exponents can be found. The most common polynomial truncations are computed around $x=0$. However, around the minimum $\kappa$ of $u_*$ yield a better and more accurate stability matrix \cite{Morris:1994ki, Aoki:1998um}.
This was indeed the main strategy used to determine the critical exponents in the works~\cite{Percacci:2015wwa, Labus:2015ska}. The ansats for the truncation is given by
\begin{equation}
u_*(x)=\lambda _0+\sum _{n=2}^{N_u} \frac{\lambda _n}{n!}(x-\kappa )^n, \quad \quad f_*(x)=\sum _{n=0}^{N_f} \frac{f_n}{n!}(x-\kappa )^n\,,
\end{equation}
where $N_u$ and $N_f$ is the order of truncation. Inserting this ansats in the fixed-point flow equations and expanding around $x=\kappa$, yields a set of coupled equations that can be solved for $\lambda_n$, $f_n$ and $\kappa$.

The critical exponents are obtained linearizing the flow equations around the polynomial ansats by eq.(\ref{anslin}) where the perturbations are given by
\begin{equation}\begin{split}
&\delta u=\delta \lambda _0+\sum _{n=2}^{N_u} \left(\frac{\delta \lambda _n}{n!}(x-\kappa )^n-\frac{  \lambda _n \delta \kappa}{(n-1)!}(x-\kappa)^{n-1}\right)\,,\\
&\delta f=\sum _{n=0}^{N_f} \left(\frac{\delta f _n}{n!}(x-\kappa )^n-\frac{f _n \delta \kappa}{(n-1)!}(x-\kappa)^{n-1}\right)\,.
\end{split}\end{equation}
The linearized flow equations then turn into a linear set of coupled equations for $\delta\lambda_n$, $\delta f_n$, $\delta \kappa$ and $\theta$. The matrix of this linear system is the stability matrix. Solving the numerical system yields the set of allowed critical exponents associated to a given scaling solution.

Compared to the polynomial truncation around $x=0$ the great precision for the critical exponents is obtained due to the possibility to set an accuracy of $10^{-64}$ in the FindRoot solutions. With the polynomial around $x=0$ the best accuracy is around $10^{-8}$.
\end{appendices}

\newpage
%\bibliography{refScal.bib}

\end{document}